\documentclass[a4paper,11pt]{article}
\pdfoutput=1 

\usepackage{jheppub2} 
\usepackage{bbm,bm,graphicx,mathtools,color,slashed,hyperref}
\usepackage{amssymb,amsmath}

\usepackage{lmodern}

\usepackage[compat=1.1.0]{tikz-feynhand}
\usetikzlibrary{patterns}
\usetikzlibrary{decorations.markings}

\graphicspath{{./Figures/}}

\newcommand{\diff}{\mathrm{d}}

\newcommand{\Diff}{{\mathcal{D}}}

\newcommand{\im}{\mathrm{i}}

\newcommand{\calN}{\mathcal{N}}

\newcommand{\calZ}{\mathcal{Z}}

\newcommand{\rmd}{\mathrm{d}}
\newcommand{\rme}{\mathrm{e}}

\newcommand{\iso}{\xrightarrow{
   \,\smash{\raisebox{-0.65ex}{\ensuremath{\scriptstyle\sim}}}\,}}
\newcommand{\Be}{\widetilde{B}_e}
\newcommand{\Bm}{B_ m}

\newcommand{\calZtH}{\mathcal{Z}_{\text{tH}}}
\newcommand{\sfS}{\mathsf{S}}
\newcommand{\sfT}{\mathsf{T}}

\DeclareMathOperator{\Tr}{Tr}
\DeclareMathOperator{\tr}{tr}

\preprint{YITP-23-72}

\title{Study of gapped phases of 4d gauge theories using temporal gauging of the $\mathbb{Z}_N$ 1-form symmetry}

\author[1]{Mendel Nguyen,}
\affiliation[1]{Department of Physics, North Carolina State University, Raleigh, NC 27607, USA}

\emailAdd{mendelnguyen@gmail.com}

\author[2]{Yuya Tanizaki,}
\affiliation[2]{Yukawa Institute for Theoretical Physics, Kyoto University,
Kitashirakawa Oiwakecho, Sakyo-ku, Kyoto 606-8502, Japan}
\emailAdd{yuya.tanizaki@yukawa.kyoto-u.ac.jp}

\author[1]{Mithat \"{U}nsal}

\emailAdd{unsal.mithat@gmail.com}

\abstract{
To study gapped phases of $4$d gauge theories, we introduce the temporal gauging of $\mathbb{Z}_N$ $1$-form symmetry in $4$d quantum field theories (QFTs), thereby defining effective $3$d QFTs with $\widetilde{\mathbb{Z}}_N\times \mathbb{Z}_N$ $1$-form symmetry. In this way, spatial fundamental Wilson and 't Hooft loops are simultaneously genuine line operators. 
Assuming a mass gap and Lorentz invariant vacuum of the $4$d QFT, the $\widetilde{\mathbb{Z}}_N\times \mathbb{Z}_N$ symmetry must be spontaneously broken to an order-$N$ subgroup $H$, and we can classify the $4$d gapped phases by specifying $H$. 
This establishes the $1$-to-$1$ correspondence between the two classification schemes for gapped phases of $4$d gauge theories: One is the conventional Wilson--'t~Hooft classification, and the other is the modern classification using the spontaneous breaking of $4$d $1$-form symmetry enriched with symmetry-protected topological states. 
}

\begin{document}
\maketitle


\section{Introduction}
\label{sec:intro}

Color confinement is one of the most remarkable phenomena in $4$d non-Abelian gauge theories, and we are continuously developing various techniques to understand its physical mechanism. Importantly, we are interested in the vacuum structure in the space of gauge theories as well as its properties for a specific theory. This motivation naturally leads us to classify the possible vacua as states of quantum phases of matter. 

The classification problem of possible gapped phases of $4$d $SU(N)$ gauge theories (with adjoint matter) has a long history, and one of the key ideas is to study the behavior of the interparticle potential for probe particles. 
We can introduce the test quark as the Wilson loop operator, and the electric charge of the test quark is characterized by the center elements of the gauge group, $\mathbb{Z}_N\subset SU(N)$. Then, Wilson proposed that confinement and Higgs phases are discriminated by studying whether the Wilson loop shows the area law or the perimeter law~\cite{Wilson:1974sk}. 
Interestingly, we can also consider magnetic particles as well as electric ones. The magnetic charges belong to $\widetilde{\mathbb{Z}}_N=\pi_1(SU(N)/\mathbb{Z}_N)$, whose elements specify possible Dirac strings, and we can describe their worldlines using 't~Hooft loops. 
The above observation leads to the Wilson--'t~Hooft classification, which says that the gapped phases are classified according to the set of deconfined dyonic lines in $\widetilde{\mathbb{Z}}_N\times \mathbb{Z}_N$~\cite{tHooft:1977nqb, tHooft:1979rtg, tHooft:1981bkw}.

In the modern perspective of generalized global symmetry in quantum field theories (QFTs), this Wilson--'t~Hooft classification is a bit mysterious. 
When we consider a $4$d gauge theory with (generalized) locality, we need to specify the global structure of the gauge group, such as $SU(N)$ vs. $SU(N)/\mathbb{Z}_N$. 
Once the global structure is specified, we cannot have both Wilson and 't~Hooft loops as genuine line operators~\cite{Aharony:2013hda, Kapustin:2014gua}. 
Only $N$ of $N^2$ dyonic lines are genuine line operators and they have to be mutually local. The other lines are non-genuine and live on the boundaries of topological surface operators, which explains the Wilson--'t~Hooft commutation relation kinematically. 
These mutually-local dyonic lines specify an order-$N$ group $G\,(\subset \widetilde{\mathbb{Z}}_N\times \mathbb{Z}_N)$, and the theory has a $G$ $1$-form symmetry~\cite{Gaiotto:2014kfa}. 
It should be noted that so far we have \textit{not} discussed the dynamics of gauge theories at all in this paragraph; everything is just about the definition of genuine line operators even though the order-$N$ subgroup of $\widetilde{\mathbb{Z}}_N\times \mathbb{Z}_N$ appears similarly as in the Wilson--'t~Hooft classification of gapped phases. 

In this paper, let us always choose the global structure of the gauge group to be $SU(N)$. Then the $1$-form symmetry is denoted by $\mathbb{Z}_N^{[1]}$, which measures the $\mathbb{Z}_N$ electric charge of the Wilson loop. 
Since the 't~Hooft lines are not genuine line operators, we do not have the magnetic counterpart of the $1$-form symmetry that measures $\widetilde{\mathbb{Z}}_N$. 
This situation would naturally raise the question of why we need the whole set of dyonic lines to characterize gapped phases in the Wilson--'t~Hooft classification. 
Here, we wish to answer this question and make a clear connection between the Wilson--'t~Hooft classification and the classification via the $1$-form symmetry.

To achieve this goal, we introduce ``temporal gauging'' of the $1$-form symmetry and apply this technique to produce $3$d QFTs with $\widetilde{\mathbb{Z}}_N\times \mathbb{Z}_N$ $1$-form symmetry out of $4$d QFTs with $\mathbb{Z}_N$ $1$-form symmetry. 
We study the partition function of these $3$d QFTs with $\widetilde{\mathbb{Z}}_N^{[1]}\times \mathbb{Z}_N^{[1]}$ in the presence of the background gauge fields, which we call the 't~Hooft partition function as it was first introduced by 't~Hooft in Ref.~\cite{tHooft:1979rtg}. 
Let us emphasize that the temporal gauging is reversible, so the 't~Hooft partition function carries the same amount of information as the $4$d partition function. 
The 't~Hooft partition function turns out to be strongly constrained by the $4$d Lorentz invariance of the original theory, and this is exactly the setup that justifies the Wilson--'t~Hooft classification. 
We show in Sec.~\ref{sec:temporalgauging} that the classification of the $4$d gapped phases according to the spontaneous breaking of the $4$d $1$-form symmetry enriched with symmetry-protected topological (SPT) states is in $1$-to-$1$ correspondence with the Wilson--'t~Hooft classification via the temporal gauging operation:
\begin{align}
    & (\mathbb{Z}_N^{[1]})_{4\rmd}\xrightarrow{\mathrm{SSB}} (\mathbb{Z}_n^{[1]})_{4\rmd} \text{ enriched with the $\mathbb{Z}_n^{[1]}$ level-$k$ SPT state}\notag\\
    \xLeftrightarrow{1:1}\quad&
    \widetilde{\mathbb{Z}}_N\times\mathbb{Z}_N\xrightarrow{\mathrm{SSB}} H=\{x(n,0)+y(k,-N/n)\in \widetilde{\mathbb{Z}}_N\times \mathbb{Z}_N\}. 
    \label{eq:equivalence_tHooft_1form}
\end{align}
The left-hand-side describes the characterization of the gapped phases in $4$d QFT language and the right-hand-side describes it after performing the temporal gauging, and these two are shown to be completely equivalent.

In Sec.~\ref{sec:anomaly}, we discuss the situation where the $4$d $1$-form symmetry has a mixed 't~Hooft anomaly. 
We shall see that the anomaly relation in $4$d is translated into the higher-group structure of $3$d QFTs after the temporal gauging. 
We can reproduce the anomaly matching constraint by combining the higher-group structure with the symmetry breaking, $\widetilde{\mathbb{Z}}_N^{[1]}\times \mathbb{Z}_N^{[1]}\xrightarrow{\mathrm{SSB}}H^{[1]}$, while the higher-group structure itself is not sufficient to reach this conclusion. 
We then introduce the $\sfS$ and $\sfT$ operations to study the connection between different gapped phases in Sec.~\ref{sec:SToperation}. 
These operations generate an $SL(2,\mathbb{Z})$ action on the space of $4$d QFTs with $\mathbb{Z}_N^{[1]}$ symmetry, and these operations give automorphisms on $\widetilde{\mathbb{Z}}_N\times \mathbb{Z}_N$ that relate different order-$N$ subgroups $H_1\iso H_2$. 
We apply it to the $\mathcal{N}=1^*$ supersymmetric Yang--Mills theory and study its rich vacuum structure from this viewpoint.

\section{Temporal gauging of \texorpdfstring{$1$}{1}-form symmetry and 't~Hooft partition function}
\label{sec:temporalgauging}

Throughout this paper, we will analyze the gapped phases of $4$d QFTs with $\mathbb{Z}_N$ $1$-form symmetry, which is denoted by $\mathbb{Z}_N^{[1]}$. 
For this purpose, we introduce the background gauge field $B_{4\rmd}$ for the $1$-form symmetry and study properties of the partition function
\begin{equation}
    \calZ[B_{4\rmd}]. 
\end{equation}
This partition function is defined on any general $4$-dimensional Riemannian manifold $M_4$. 
In the following, we especially pay attention to the case
\begin{equation}
    M_4=M_3\times S^1.
\end{equation}
We refer to this $S^1$ as the temporal direction, and denote its coordinate by $\tau\sim \tau+L$. 
As we are still interested in the four-dimensional dynamics, we basically assume that the size $L$ of $S^1$ is sufficiently large and the phase is smoothly connected to the ground states. 
In the following, we choose a spin structure for $M_3\times S^1$. 

By regarding the size of $M_3$ to be much larger than that of $S^1$, we can pretend that we are dealing with $3$d QFTs. 
Then, the $4$d $\mathbb{Z}_N^{[1]}$ symmetry splits into~\cite{Gaiotto:2014kfa} 
\begin{equation}
    (\mathbb{Z}_N^{[1]})_{4\rmd}\Longrightarrow 
    (\mathbb{Z}_N^{[1]})_{3\rmd}\times (\mathbb{Z}_N^{[0]})_{3\rmd}. 
\end{equation}
Let $\Bm$ and $A$ denote the background gauge fields for $(\mathbb{Z}_N^{[1]})_{3\rmd}$ and $(\mathbb{Z}_N^{[0]})_{3\rmd}$, respectively. Then they can be related to the $4$d background gauge fields $B_{4\rmd}$ as~\cite{Gaiotto:2017yup, Shimizu:2017asf, Tanizaki:2017qhf, Tanizaki:2017mtm}
\begin{equation}
    B_{4\rmd}=\Bm+A\wedge \frac{\diff \tau}{L}. 
\end{equation}
Here, $\Bm$ does not have the temporal component, and we sometimes call it the magnetic flux. 
The temporal-spatial component is expressed by the $1$-form gauge field $A$. 

We define the temporal gauging by the path integral in terms of $A$:\footnote{We follow the convention that the background fields are denoted with upper case and the dynamical ones are with lower case. When we promote the background gauge fields to the dynamical ones, we change their letters to the corresponding lower case letters, accordingly. } 
\begin{equation}
    \calZtH[\Be, \Bm]= \int \Diff a \,\exp\biggl(\frac{2\pi \im}{N}\int_{M_3} \Be\cup a\biggr) \,\calZ[\Bm+a\wedge (\diff \tau/L)]. 
    \label{eq:tHooftZ}
\end{equation}
Here, $\Be$ is the background gauge field for the $\mathbb{Z}_N$ $1$-form symmetry, $\widetilde{\mathbb{Z}}_{N}^{[1]}$, dual to the original $\mathbb{Z}_N^{[0]}$ symmetry. 
Regarding $\calZtH$ as the partition function of the $3$d QFT defined on $M_3$, it enjoys the $\widetilde{\mathbb{Z}}_N^{[1]}\times \mathbb{Z}_N^{[1]}$ symmetry and $(\Be, \Bm)$ is the corresponding background $2$-form gauge field. 
As this partition function~\eqref{eq:tHooftZ} was first introduced by 't~Hooft in Ref.~\cite{tHooft:1979rtg} for the case of $M_4=T^4$, we shall refer to it as the 't~Hooft partition function. 

As this theory enjoys the $\widetilde{\mathbb{Z}}_N\times \mathbb{Z}_N$ $1$-form symmetry, we must have the corresponding line operators. 
Since the $\mathbb{Z}_N^{[1]}$ symmetry is the $1$-form symmetry in the original $4$d theory, let us refer to the corresponding operator as the Wilson loop, $W_{(0,1)}(C)$. 
It is then natural to refer to the charged object of $\widetilde{\mathbb{Z}}_N^{[1]}$ as the 't~Hooft loop, $W_{(1,0)}(C)$. In general, the dyonic loop operator (with the magnetic charge $m\in\mathbb{Z}_N$ and the electric charge $e\in\mathbb{Z}_N$) is denoted as 
\begin{equation}
    W_{(m,e)}(C)\,\rme^{\im \int_{D}(m \Be + e \Bm)}
\end{equation}
with some $\partial D=C$ in the presence of the background gauge fields.\footnote{We note that these are genuine line operators in the effective $3$d QFT on $M_3$ as the surface dependence appears only when we turn on background $2$-form gauge fields. } 

Here, we need to emphasize that the 't~Hooft partition function~\eqref{eq:tHooftZ} is introduced to understand the possible phases of $4$d gauge theories, while $\Be$ and $\Bm$ there are $3$d $\mathbb{Z}_N$ $2$-form gauge fields. 
Even though $\calZ[B_{4\rmd}]$ is covariant under Lorentz transformations, the temporal-gauging procedure does not respect it, and thus $\calZtH$ may seem at first sight to be less useful compared with the original one $\calZ[B_{4\rmd}]$ for studying the $4$d dynamics. 
Let us point out, however, that the temporal gauging is a reversible operation, and thus $\calZ[B_{4\rmd}]$ and $\calZtH[\Be, \Bm]$ should carry the same amount of information. 
Moreover, it turns out in the following that $\calZtH[\Be,\Bm]$ provides a convenient tool for the classification of gapped phases, and the physical meaning of each phase also becomes quite transparent. 

\subsection{Positivity of the 't~Hooft partition function \texorpdfstring{$\calZtH[\Be,\Bm]$}{Zth[Be,Bm]}}
\label{sec:positivity}

An important property of the 't~Hooft partition function is its semi-positivity, and one can easily show it using reflection positivity.\footnote{We assume that the $4$d theory is unitary and thus its path integral satisfies reflection positivity. } 
Regard $S^1$ as the temporal direction. We pick the antipodal points $\{\tau=0,L/2\}\subset S^1$ and choose $M_3\times \{0, L/2\}\subset M_4$ as the reflection plane for the Osterwalder--Schrader reflection. 

Using the $1$-form gauge invariance, or the topological nature of the codim-$2$ defects, we may set the specific alignment of the discrete gauge fields. 
For the spatial part $\Bm$, we require that $\Bm$ does not depend on $\tau$ at all so that $\Bm$ is invariant under the Osterwalder--Schrader reflection. 
To discuss the temporal gauge field $a\wedge (\diff \tau/L)$, we note the following trivial identity, 
\begin{equation}
    \int \Diff a\, F[a]=\int \Diff a_1 \Diff a_2\, F[a_1-a_2], 
\end{equation}
which holds for any functionals $F[a]$. Using this identity, we find 
\begin{equation}
    \calZtH[\Be, \Bm]=\int \Diff a_1 \Diff a_2\, \rme^{\frac{2\pi \im}{N}\int_{M_3} \Be \cup (a_1-a_2)}  \calZ\biggl[\Bm + a_1\wedge \frac{\diff \tau}{L}-a_2 \wedge \frac{\diff \tau}{L}\biggr]. 
\end{equation}
Thus, the defects for the temporal directions can be doubled, and these two defects can be put on arbitrary locations due to their topological nature. 
By a suitable choice, they can be related by the Osterwalder--Schrader reflection. 
Then, the reflection positivity ensures that 
\begin{equation}
    \calZtH[\Be, \Bm]\ge 0
    \label{eq:positivity}
\end{equation}
for any $\Be, \Bm\in H^2(M_3;\mathbb{Z}_N)$. 

We note that this positivity is achieved by the temporal gauging procedure. 
Indeed, the ordinary partition function, $\calZ[B_{4\rmd}]$, can take complex values in general in the presence of the background gauge fields, and it often provides us important information on the quantum phases of matter. 
In general, temporal components of the gauge field flip their sign under the Osterwalder--Schrader reflection, so this complex phase is consistent with reflection positivity. 
In the case of the 't~Hooft partition function, the positivity argument works as we sum up all the possible gauge fields having temporal components (note that $\Be$ only has purely spatial components), and we find \eqref{eq:positivity}. 

The physical meaning of the positivity~\eqref{eq:positivity} becomes more transparent if we consider it in the operator formalism~\cite{tHooft:1979rtg}. 
Let $\mathcal{H}$ be the Hilbert space when we quantize the theory on $M_3$, and let $\hat{H}[\Bm]$ be the Hamiltonian operator that contains the magnetic flux $\Bm$. 
We can further define the projection operator onto the electric flux sector $\hat{P}[\Be]$, which satisfies $\hat{P}[\Be]^\dagger=\hat{P}[\Be]$, $\hat{P}[\Be]^2 = \hat{P}[\Be]$ and $\sum_{\Be} \hat{P}[\Be]=\bm{1}_{\mathcal{H}}$. 
Then the 't~Hooft partition function can be written as 
\begin{equation}
    \calZtH[\Be,\Bm]=\Tr_{\mathcal{H}}\Bigl[\hat{P}[\Be]\,\rme^{-L \hat{H}[\Bm]}\Bigr], 
\end{equation}
and then its positivity is quite manifest.  

\subsection{\texorpdfstring{$\calZtH[\Be,\Bm]$}{ZtH[Be,Bm]} for gapped phases and constraints from Lorentz invariance}

In Ref.~\cite{tHooft:1979rtg}, 't~Hooft found the ``duality equation'' for $\calZtH[\Be, \Bm]$ by considering a discrete rotation of the torus $M_4=T^4$. By assuming that the system is gapped, the duality equation implies that $\widetilde{\mathbb{Z}}_N^{[1]}\times \mathbb{Z}_N^{[1]}$ should be spontaneously broken to an order-$N$ subgroup, i.e., 
\begin{equation}
    \widetilde{\mathbb{Z}}_N^{[1]}\times \mathbb{Z}_N^{[1]} \xrightarrow{\mathrm{SSB}} H^{[1]},
\end{equation}
where the unbroken symmetry $H$ has order $N$, $|H|=N$, and satisfies\footnote{We note that when $H$ is an order-$N$ subgroup of $\widetilde{\mathbb{Z}}_N\times \mathbb{Z}_N$, then the mutual locality condition~\eqref{eq:mutuallocality} is automatically satisfied. 
To see this, let us assume (to derive a contradiction) that the mutual locality is violated, so that there are $(x_1, y_1), (x_2,y_2) \in H$ such that $M\equiv x_1 y_2 - x_2 y_1 \not = 0 \bmod N$. Without loss of generality, this $M \,(<N)$ can be taken to be a positive divisor of $N$. 
Then, the subgroup $\{a (x_1, y_1)+b(x_2, y_2)\} \subset H$ has order $N^2/M>N$, and this contradicts with $|H|=N$. }
\begin{equation}
    \forall (x_1, y_1), (x_2, y_2)\in H, \quad \langle (x_1,y_1), (x_2,y_2)\rangle \equiv x_1 y_2 - x_2 y_1 =0 \bmod N. 
    \label{eq:mutuallocality}
\end{equation} 
This condition is referred to as the mutual locality condition. 

In order to see how such a constraint on gapped phases arises, let us consider what would happen if $\widetilde{\mathbb{Z}}_N^{[1]}\times \mathbb{Z}_N^{[1]}$ were not broken at all. By gauging $\Be$, we can undo the temporal gauging procedure and this gives the delta-functional constraint on $A$ for $(\mathbb{Z}_N^{[0]})_{3\rmd}$. This implies that $(\mathbb{Z}_N^{[0]})_{3\rmd}$ is spontaneously broken, while $(\mathbb{Z}_N^{[1]})_{3\rmd}$ is unbroken by assumption. 
Since both of these symmetries arise from $(\mathbb{Z}_N^{[1]})_{4\rmd}$, this option obviously violates the $4$d Lorentz invariance. 
Similarly, if we assume $\widetilde{\mathbb{Z}}_N^{[1]}\times \mathbb{Z}_N^{[1]}$ were completely broken, we find that $(\mathbb{Z}_N^{[0]})_{3\rmd}$ is unbroken while $(\mathbb{Z}_N^{[1]})_{3\rmd}$ is broken, and again Lorentz invariance is violated. 
These quick observations already tell us that Lorentz invariance puts severe constraints and requires the correct amount of symmetry breaking for $\widetilde{\mathbb{Z}}_N^{[1]}\times \mathbb{Z}_N^{[1]}$, and we can actually find that it is broken down to an exactly order-$N$ subgroup (with mutual locality) when assuming a mass gap (and also one technical assumption).\footnote{As this consequence is very similar to that of the anomaly-matching constraint, one might wonder if this can be understood from the mixed anomaly for $\widetilde{\mathbb{Z}}_N\times \mathbb{Z}_N$ $1$-form symmetry. However, this is not the case, and we emphasize that the $4$d Lorentz invariance of the original theory plays a pivotal role here. } 

We shall give a review of the original argument by 't~Hooft in Appendix~\ref{sec:tHooftoriginal} to be self-contained. 
Here, instead, let us perform explicit calculations of the $4$d partition function $\calZ[B_{4\rmd}]$ and $\calZtH[\Be,\Bm]$ for gapped phases. 
We here assume that the 4d $\mathbb{Z}_N$ $1$-form symmetry is spontaneously broken to a subgroup, 
\begin{equation}
    (\mathbb{Z}_N^{[1]})_{4\rmd}\xrightarrow{\mathrm{SSB}}(\mathbb{Z}_n^{[1]})_{4\rmd}, 
\end{equation}
where $n$ is a positive divisor of $N$, and the vacuum state further acquires a nontrivial SPT phase for the unbroken $(\mathbb{Z}_n^{[1]})_{4\rmd}$ symmetry. 
The low-energy theory becomes $\mathbb{Z}_{N/n}$ topological field theory, and the partition function can be modeled as\footnote{We may consider more general $2$-group gauge theories as possible models (see, e.g., Refs.~\cite{Gukov:2013zka, Kapustin:2013uxa, Kapustin:2013qsa, Lan:2018vjb, Johnson-Freyd:2020usu,  Thorngren:2020aph}). Here, let us restrict our attention to these simplest possibilities. } 
\begin{align}
    \calZ[B_{4\rmd}]&=
    \frac{|H^0(M_4;\mathbb{Z}_{N/n})|}{|H^1(M_4;\mathbb{Z}_{N/n})|}\sum_{b\in H^2(M_4;\mathbb{Z}_{N/n})} \exp\biggl(\frac{2\pi \im}{N/n}\int b\cup  B_{4\rmd}\biggr)\nonumber\\
    &\quad \times \exp\biggl(\frac{2\pi \im k}{n}\int \frac{1}{2}P_2\biggl(\frac{B_{4\rmd}}{N/n}\biggr)\biggr), 
    \label{eq:Z_modelcomputation}
\end{align}
where $P_2(B)=B\cup B+B\cup_1 \diff B$ is the Pontryagin square.\footnote{As long as working on torsion-free $4$-manifolds, we can always take an integral lift of the discrete gauge field $B$ and $P_2(B)$ can be simply thought of as $B\cup B$ by identifying $B$ with one of the integral lifts. We shall use this property throughout this paper to simplify computations.} 
The $b$ field refers to the discrete $\mathbb{Z}_{N/n}$ $2$-form gauge field (\textit{not} a $\mathbb{Z}_N$ gauge field) for the topological field theory, and its path integral gives the delta-functional constraint on $B_{4\rmd}$ so that 
\begin{equation}
    \int_{\Sigma} B_{4\rmd}\in \frac{N}{n}\mathbb{Z}
\end{equation}
for any closed $2$-cycle $\Sigma$. As $\int B_{4\rmd}$ is well-defined $\bmod\, N$, we can regard $B_{4\rmd}/(N/n)$ as the $\mathbb{Z}_n$ $2$-form gauge field, and the second line on the right-hand-side of \eqref{eq:Z_modelcomputation} describes the level-$k$ SPT action for this unbroken $\mathbb{Z}_n^{[1]}$ symmetry with $k\sim k+n$ (given a spin structure). 

Let us compute the 't~Hooft partition function for \eqref{eq:Z_modelcomputation}, which is given by 
\begin{align}
    \calZtH[\Be, \Bm]&=\frac{1}{|H^0(M_3;\mathbb{Z}_N)|}\sum_{a\in H^1(M_3;\mathbb{Z}_N)}\exp\biggl(\frac{2\pi \im}{N}\int \Be \cup a\biggr)\notag\\
    &\quad \times \frac{|H^0(M_3\times S^1;\mathbb{Z}_{N/n})|}{|H^1(M_3\times S^1;\mathbb{Z}_{N/n})|}\sum_{b_m\in H^2(M_3;\mathbb{Z}_{N/n})}\sum_{a'\in H^1(M_3;\mathbb{Z}_{N/n})}\notag\\
    &\quad \times \exp\biggl(\frac{2\pi \im}{N/n}\int (b_m \cup a + \Bm \cup a')+\frac{2\pi \im}{n}\frac{k}{(N/n)^2}\int \Bm \cup a\biggr). 
\end{align}
We note that $|H^1(M_3\times S^1;\mathbb{Z}_{N/n})|=(N/n)^{\beta_1(M_3)+1}$, so $\frac{|H^0(M_3\times S^1;\mathbb{Z}_{N/n})|}{|H^1(M_3\times S^1;\mathbb{Z}_{N/n})|}=\frac{1}{(N/n)^{\beta_1(M_3)}}$. 
Here, $\beta_i(M)=\mathrm{rank}\, H^i(M;\mathbb{Z})$ is the $i$-th Betti number. 
The summation over $a'$ gives the delta-functional constraint on $B_m$, so we separate it from other path integrals:
\begin{align}
    \calZtH[\Be,\Bm]&= \frac{1}{(N/n)^{\beta_1(M_3)}}\sum_{a'\in H^1(M_3;\mathbb{Z}_{N/n})}\exp\biggl(\frac{2\pi \im}{N/n}\int a' \cup \Bm\biggr)\notag\\
    &\quad \times\frac{1}{N}\sum_{a\in H^1(M_3;\mathbb{Z}_N)}\sum_{b_m\in H^2(M_3;\mathbb{Z}_{N/n})}\exp\biggl(\frac{2\pi \im}{N}\int a \cup \biggl(n\, b_m + \Be + k\frac{n}{N}\Bm \biggr)\biggr)  
    \notag\\
    &= \frac{1}{(N/n)^{\beta_1(M_3)}} (N/n)^{\beta_1(M_3)} \delta_N[n \Bm]  \times \frac{1}{N} N^{\beta_1(M_3)}\delta_N\biggl[\frac{N}{n}\biggl(\Be + k \frac{n}{N}\Bm\biggr)\biggr] 
    \notag\\
    &=N^{\beta_1(M_3)-1}\, \delta_N\biggl[n \Bm, \,\frac{N}{n}\Be + k \Bm\biggr]. 
    \label{eq:tHooftZ_GappedPhases}
\end{align}
Here, $\delta_N[B]$ is the delta functional that gives $1$ when $\int_{\Sigma}B \in N\mathbb{Z}$ for every closed cycle $\Sigma$ and gives $0$ otherwise. 
This shows that the deconfined lines are generated by 
\begin{equation}
    W_{(0,n)}(C)\,\rme^{\im \int_{D} n\Bm}, \quad 
    W_{(N/n,k)}(C)\,\rme^{\im \int_{D}((N/n)\Be+ k \Bm)},
\end{equation}
with $\partial D=C$, 
and thus the unbroken subgroup $H$ is given by 
\begin{equation}
    H=\left\{x\Big(n,0\Big)+y\Big(k,-\frac{N}{n}\Big)\right\} \subset \widetilde{\mathbb{Z}}_N\times \mathbb{Z}_N,
    \label{eq:unbrokensubgroup}
\end{equation}
and we can readily confirm that the mutual locality condition~\eqref{eq:mutuallocality} is satisfied. 
We can also check that $|H|=N$ and moreover that every order $N$ subgroup of $\widetilde{\mathbb{Z}}_N\times \mathbb{Z}_N$ appears in this way.\footnote{To see that every order $N$ subgroup of $\mathbb{Z}_N \times \mathbb{Z}_N$ appears, note that any such subgroup $K$ arises from an index $N$ sublattice $L$ of $\mathbb Z \times \mathbb Z$ containing $N \mathbb{Z} \times N \mathbb{Z}$ such that $K = L / (N \mathbb Z \times N\mathbb Z)$. Then, since $(N,0)$ and $(0,N)$ are linearly independent vectors in $L$, a theorem on lattices implies that we can find a basis $u,v$ of $L$ of the form $u \equiv \frac{1}{q}(N,0), v \equiv \frac{k}{N} (N,0) + \frac{1}{n} (0,N)$, with $q,n$ positive divisors of $N$ and $k$ an integer. The condition that $L$ have index $N$ in $\mathbb Z \times \mathbb Z$ then implies that $N^2/qn = \langle u,v \rangle = N$, i.e., that $N/q = n$. Hence, $K$ is precisely of the form \eqref{eq:unbrokensubgroup}.}

In the above discussion, we start from the $4$d partition function~\eqref{eq:Z_modelcomputation} and derive \eqref{eq:tHooftZ_GappedPhases} by the temporal gauging, but we can reverse the logic to reproduce \eqref{eq:Z_modelcomputation} by performing the path integral of $\Be$ of the 't~Hooft partition function~\eqref{eq:tHooftZ_GappedPhases}, which achieves the equivalence mentioned in~\eqref{eq:equivalence_tHooft_1form}, and let us recapitulate it here:
\begin{align}
    & (\mathbb{Z}_N^{[1]})_{4\rmd}\xrightarrow{\mathrm{SSB}} (\mathbb{Z}_n^{[1]})_{4\rmd} \text{ enriched with the $\mathbb{Z}_n^{[1]}$ level-$k$ SPT state}\notag\\
    \xLeftrightarrow{1:1}\quad&
    \widetilde{\mathbb{Z}}_N\times\mathbb{Z}_N\xrightarrow{\mathrm{SSB}} H=\{x(n,0)+y(k,-N/n)\in \widetilde{\mathbb{Z}}_N\times \mathbb{Z}_N\}. 
\end{align}
Therefore, the order-$N$ subgroup $H$ of $\widetilde{\mathbb{Z}}_N\times \mathbb{Z}_N$ correctly characterizes the gapped phases of $4$d QFTs with $\mathbb{Z}_N^{[1]}$ symmetry: Vacuum states with different $H$ are distinguished as quantum phases. 

\subsection{Example: Lattice \texorpdfstring{$SU(N)$}{SU(N)} Yang--Mills theory at strong coupling}

It would be useful to compute $\calZ[B_{4\rmd}]$ and $\calZtH[\Be, \Bm]$ in some microscopically solvable model for concrete understanding of their behaviors. Here, let us consider the strong-coupling expansion of the lattice $SU(N)$ gauge theory with the Wilson action. 

The Wilson action with the $\mathbb{Z}_N$ two-form gauge field is given by 
\begin{equation}
    S_W[U_\ell, B_p]=-\frac{1}{2g^2} \sum_{p}\tr\Bigl(\rme^{-\frac{2\pi \im }{N} B_p}U_p + \rme^{\frac{2\pi \im}{N}B_p}U_p^\dagger\Bigr), 
\end{equation}
where $U_\ell$ denotes the $SU(N)$-valued link variable, $U_p=\mathcal{P}\prod_{\ell\in p} U_\ell$ is the path-ordered products along the plaquette $p$, and $B_p$ denotes the $\mathbb{Z}_N$-valued plaquette variable, which is identified with $B_{4\rmd}$. The partition function is given by 
\begin{equation}
    \calZ[B_{4\rmd}]=\int \Diff U_\ell \exp\Bigl(-S_W[U_\ell, B_p]\Bigr). 
\end{equation}
We expand this partition function in terms of $1/g^2$ in the strong-coupling expansion by using formulas of Haar integration, such as $\int \diff U (U)_{i_1 i_2} (U^\dagger)_{j_1 j_2}=\frac{1}{N}\delta_{i_1 j_2} \delta_{i_2 j_1}$. 
Let us then expand the path-integral weight up to the $O(1/g^2)$ term for each plaquette,  
\begin{equation}
    \exp(-S_W)\simeq \prod_{p}\left(1+\frac{1}{2g^2} \tr (\rme^{-\frac{2\pi \im }{N} B_p}U_p + \rme^{\frac{2\pi \im}{N}B_p}U_p^\dagger)\right).
\end{equation} 
Then the partition function can be represented as a sum over closed surfaces, 
\begin{equation}
    \calZ[B_{4\rmd}]\simeq \sum_{\Sigma\colon \text{closed surface}} N^{\chi(\Sigma)}\left(\frac{1}{2Ng^2}\right)^{\mathrm{Area}(\Sigma)}\rme^{-\frac{2\pi \im}{N} \int_{\Sigma}B_{4\rmd}}. 
    \label{eq:strongcoupling}
\end{equation}
We can think of this expression as the sum over the worldsheets of confining strings with the string tension $\sigma=\ln (2Ng^2)$ in lattice units. 
When $\Sigma$ is a contractible closed surface, we have $\int_{\Sigma} B_{4\rmd}=0 \bmod N$. Thus, the nontrivial $B_{4\rmd}$ dependence appears only if the confining-string worldsheet wraps around nontrivial $2$-cycles, and such processes are exponentially suppressed:
\begin{equation}
    \calZ[B_{4\rmd}]-\calZ[0]\simeq O(\rme^{-\sigma L^2})\xrightarrow{L\to \infty}0,
\end{equation}
where $L$ is the length of $T^4$. 
This shows that, in the infinite-volume limit, we can regard $\calZ[B_{4\rmd}]\to 1$, which corresponds to $n=N$ and $k=0$ in \eqref{eq:Z_modelcomputation}.

Now, let us perform the temporal gauging of \eqref{eq:strongcoupling} to find $\calZtH[\Be,\Bm]$. 
As we have found that the $B_{4\rmd}$ dependence of $\mathcal{Z}[B_{4\rmd}]$ is exponentially small, its Fourier transform localizes to $\Be=0$ and we get $\calZtH[\Be, \Bm]\propto \delta_N[\Be]$. 
More precisely, for $\Be=0$,
\begin{align}
    \calZtH[0,\Bm]&=\int \Diff a\, \calZ[B_m+a\wedge \diff \tau/L] \notag\\
    &\simeq \calZ[0]+O(\rme^{-\sigma L^2}) \xrightarrow{L\to \infty} \calZ[0]. 
\end{align}
On the other hand, if we take $\int_{(T^2)_{12}}\Be=1$ as an example of $\Be\not = 0$, then $\exp(\im \int \Be\cup a)=\exp(\im \int_{S^1}a_3 \diff x^3)$. 
To cancel this phase in the summation of $a$, the confining-string worldsheet should wrap once around the $3$-$4$ cycle, and we get 
\begin{align}
    \calZtH[\Be(\not=0),\Bm]&=\int \Diff a \,\rme^{\frac{2\pi \im}{N}\int a_3 \diff x^3}\calZ[\Bm+a\wedge \diff \tau/L] \notag\\ 
    &\simeq O(\rme^{-\sigma L^2}) \xrightarrow{L\to \infty} 0. 
\end{align}
We actually find $\calZtH[\Be, \Bm]=\calZ[0]\,\delta_N[\Be]$ neglecting the exponentially small contributions as $L\to \infty$, and the unbroken order-$N$ subgroup is $H=\{0\}\times \mathbb{Z}_N\subset \widetilde{\mathbb{Z}}_N\times \mathbb{Z}_N$.

\section{Anomaly matching and the higher-group structure}
\label{sec:anomaly}

In general, global symmetry in QFTs may have an 't~Hooft anomaly, which is an obstruction to the promotion of the global symmetry to local gauge redundancy. 
The 't~Hooft anomaly is invariant under any local and symmetric deformations of QFTs, and thus the low-energy effective theory is strongly constrained as it must reproduce the anomaly computed in ultraviolet. 
In this section, we shall discuss the structure of the 't~Hooft partition function $\calZtH[\Be,\Bm]$ when the original $4$d $\mathbb{Z}_N^{[1]}$ symmetry has a mixed 't~Hooft anomaly. 

{\bf Pure Yang--Mills theory:}
As an example, let us consider the generalized anomaly, or global inconsistency, of pure $SU(N)$ Yang--Mills theory. 
The $4$d Yang--Mills partition function $\calZ^{\mathrm{YM}}_{\theta}$ has an 't~Hooft anomaly involving the $\theta$ periodicity~\cite{Gaiotto:2017yup, Tanizaki:2017bam, Karasik:2019bxn, Cordova:2019uob} and we can detect it by introducing the background $\mathbb{Z}_N$ two-form gauge field $B_{4\rmd}$: 
\begin{equation}
    \calZ^{\mathrm{YM}}_{\theta+2\pi}[B_{4\rmd}]=\exp\biggl(\frac{2\pi \im}{N}\int \frac{1}{2}P_2(B_{4\rmd})\biggr)\calZ^{\mathrm{YM}}_{\theta}[B_{4\rmd}].
    \label{eq:YM_anomaly}
\end{equation}
To satisfy the anomaly matching condition in the confined phase, the level crossing of the ground state is mandatory, as the two confined states at $\theta$ and $\theta+2\pi$ are distinct as $4$d SPT states with $\mathbb{Z}_N^{[1]}$ symmetry. 

Let us interpret this result using the 't~Hooft partition function:
\begin{equation}
    \calZ_{\mathrm{tH},\theta}^{\mathrm{YM}}[\Be,\Bm]=\int \Diff a \exp\biggl(\frac{2\pi\im}{N}\int \Be \cup a\biggr) \calZ_{\theta}^{\mathrm{YM}}[\Bm+a\wedge \diff \tau/L]. 
\end{equation}
By performing the temporal gauging on both sides of \eqref{eq:YM_anomaly}, we find that 
\begin{align}
    \calZ_{\mathrm{tH},\theta+2\pi}^{\mathrm{YM}}[\Be,\Bm]
    &=\int \Diff a\, \rme^{\frac{2\pi\im}{N}\int \Be a} \calZ^{\mathrm{YM}}_{\theta+2\pi}[\Bm+a\wedge \diff \tau/L] \notag\\
    &=\int \Diff a\, \rme^{\frac{2\pi\im}{N}\int \Be a} \rme^{\frac{2\pi \im}{N}\int \Bm a} \calZ^{\mathrm{YM}}_{\theta+2\pi}[\Bm+a\wedge \diff \tau/L] \notag\\
    &=\calZ_{\mathrm{tH},\theta}^{\mathrm{YM}}[\Be+\Bm,\Bm]. 
    \label{eq:highergroup_YM}
\end{align}
This is nothing but the Witten effect~\cite{Witten:1979ey}, which claims that the purely magnetic line at $\theta+2\pi$ is equivalent to the dyonic line at $\theta$.
Equivalently, the $(-1)$-form transformation, $\theta \to \theta+2\pi$, induces the nontrivial action on the $1$-form symmetries, $\Be\to \Be+\Bm$, which is an example of the higher-group structure~\cite{Sharpe:2015mja, Tachikawa:2017gyf, Cordova:2018cvg, Tanizaki:2019rbk, Hidaka:2020iaz, Hidaka:2020izy}. 

We note that the Witten effect, or the higher-group structure, itself does not give nontrivial constraints, in constrast to the 't~Hooft anomaly. The trivial state, $\calZ_{\mathrm{tH},\theta}^{\mathrm{YM}}[\Be,\Bm]=1$, is consistent with the transformation, $\Be\to \Be+\Bm$, and this state is indeed realized by the high-temperature Yang--Mills theory at any value of $\theta$. 
The anomaly matching constraint is reproduced by considering the $4$d Lorentz invariance. 
As discussed in Sec.~\ref{sec:temporalgauging}, the gapped state with $4$d Lorentz invariance must have the symmetry breaking $\widetilde{\mathbb{Z}}_N^{[1]}\times \mathbb{Z}_N^{[1]}\xrightarrow{\mathrm{SSB}}H^{[1]}$, and thus we can set the 't~Hooft partition function for a given $\theta$ to be
\begin{equation}
    \calZ_{\mathrm{tH},\theta}^{\mathrm{YM}}[\Be,\Bm]=\delta_N[n\Bm]\delta_N[(N/n)\Be+k\Bm]. 
\end{equation}
Dialing the $\theta$ parameter, $\theta\to \theta+2\pi$, we should obtain 
\begin{equation}
    \calZ_{\mathrm{tH},\theta+2\pi}^{\mathrm{YM}}[\Be,\Bm]=\delta_N[n\Bm]\delta_N[(N/n)\Be+(k+(N/n))\Bm].
\end{equation}
As $k\sim k+n$, we should encounter a phase transition as a function of $\theta$ if $N/n$ is not a multiple of $n$. 
This is always the case for ordinary confinement phases, $n=N$, while the totally Higgs phase, $n=1$, does not need the phase transition in $\theta$. This reproduces the consequence of the $4$d 't~Hooft anomaly~\eqref{eq:YM_anomaly}. 

{\bf $\calN=1$ supersymmetric Yang--Mills theory:}
The higher group structure may be more evident in the $\mathcal{N}=1$ super Yang--Mills (SYM) case, where the shift of the $\theta$ angle is related to the discrete chiral symmetry $(\mathbb{Z}_{2N})_\chi$. 
Introducing the discrete chiral gauge field $A_\chi$, we find the 't~Hooft anomaly,
\begin{equation}
    \calZ^{\mathrm{SYM}}[A_\chi+\diff \lambda_\chi, B_{4\rmd}]
    = \exp\biggl(\frac{2\pi \im}{N}\int \lambda_\chi\cup \frac{1}{2}P_2(B_{4\rmd})\biggr)
    \calZ^{\mathrm{SYM}}[A_\chi, B_{4\rmd}]. 
\end{equation}
By performing a similar computation as in \eqref{eq:highergroup_YM}, this relation is translated as 
\begin{equation}
    \calZ^{\mathrm{SYM}}_{\mathrm{tH}}[A_\chi+\diff \lambda_\chi, \Be, \Bm]
    =\calZ^{\mathrm{SYM}}_{\mathrm{tH}}[A_\chi, \Be+\lambda_\chi \cup \Bm, \Bm]. 
\end{equation}
The $0$-form chiral symmetry causes the Witten effect and induces a nontrivial action on the $1$-form symmetry, $\Be\to \Be+\lambda_\chi\cup \Bm$. 

Again, the higher-group symmetry itself does not require the degeneracy of ground states, but it gives a nontrivial consequence when we further impose the $4$d Lorentz invariance. 
If we assume that the system is in a confined phase (i.e. $n=N$), then the 't~Hooft argument shows that the partition function of a given vacuum should be described by 
\begin{equation}
    \delta_N[\Be+k \Bm], 
\end{equation}
with some $k\sim k+N$. Then, the higher-group structure discussed above indicates that the discrete chiral transformation interchanges the vacuum with label $k$ to the vacuum with label $k+1$:
\begin{align}
\begin{tikzpicture}[auto,->]
\node (a) at (0,0) {$\delta_N[\Be]$};
\node (b) at (2.7,0) {$\delta_N[\Be+\Bm]$};
\node (c) at (5.2,0) {$\cdots$};
\node (d) at (8.4,0) {$\delta_N[\Be+(N-1)\Bm]$};
\draw (a) -- node {\scriptsize chiral} (b);
\draw (b) -- node {\scriptsize chiral} (c);
\draw (c) -- node {\scriptsize chiral} (d);
\draw (8.,-0.3) .. controls (6.3,-0.6) and (2.1,-0.6) .. (0.2,-0.3)
node[pos=0.5]{\scriptsize chiral};
\end{tikzpicture}
\label{eq:chiralbrokenvacua}
\end{align}
The 't~Hooft partition function of $\mathcal{N}=1$ SYM theory is then given by
\begin{equation}
    \calZtH^{\mathrm{SYM}}[\Be,\Bm]=\sum_{k=1}^{N} \delta_N[\Be+k\Bm]. 
\end{equation}
These $N$ vacua are understood as the chiral broken vacua, $(\mathbb{Z}_{2N})_\chi\xrightarrow{\mathrm{SSB}}\mathbb{Z}_2$, and the label $k$ specifies the phase of the gluino condensate, $\langle \lambda^2\rangle=\Lambda^3\rme^{2\pi \im k/N}$. 
This is exactly the vacuum structure expected from the anomaly matching condition obtained before the temporal gauging, and the same information is found via the higher-group structure combined with the constraint from $4$d Lorentz invariance.

\section{\texorpdfstring{$\sfS$}{S} and \texorpdfstring{$\sfT$}{T} operations on \texorpdfstring{$4$}{4}d QFTs with the \texorpdfstring{$\mathbb{Z}_N$ $1$}{ZN 1}-form symmetry}
\label{sec:SToperation}

Let us introduce the formal operations, $\sfS$ and $\sfT$, that act on $4$d QFTs with the $\mathbb{Z}_N$ $1$-form symmetry:
\begin{align}
    \sfS &\colon \calZ[B_{4\rmd}]\mapsto \sfS \calZ[B_{4\rmd}]\equiv\int \Diff b_{4\rmd} \,\calZ[b_{4\rmd}]\exp\biggl(\frac{2\pi \im}{N}\int B_{4\rmd}\cup b_{4\rmd}\biggr), 
    \label{eq:Strans} \\
    \sfT &\colon \calZ[B_{4\rmd}]\mapsto \sfT \calZ[B_{4\rmd}]\equiv \calZ[B_{4\rmd}]\exp\biggl(\frac{2\pi \im}{N}\int \frac{1}{2}P_2(B_{4\rmd})\biggr).
    \label{eq:Ttrans}
\end{align}
The $\sfS$ operation dynamically gauges the $\mathbb{Z}_N^{[1]}$ in $4$d spacetime, and thus the original background gauge field is promoted to the dynamical field $b_{4\rmd}$. The gauged theory acquires the dual $\mathbb{Z}_N^{[1]}$ symmetry, and we introduce the background gauge field $B_{4\rmd}$ that couples to it. 
The $\sfT$ operation just shifts the local counterterm for the background gauge field $B_{4\rmd}$.

Here, we would like to emphasize that these $\sfS$ and  $\sfT$ operations do not necessarily imply the duality/symmetry of a given $4$d QFT. 
We can always apply these operations as long as the $4$d QFTs have a $\mathbb{Z}_N$ $1$-form symmetry, and generically these operations generate different QFTs. 
One may say that $\sfS$ and $\sfT$ are morphisms in the category of $4$d QFTs with the $\mathbb{Z}_N^{[1]}$ symmetry (see Refs.~\cite{Kapustin:1999ha, Witten:2003ya} for the case of $3$d $U(1)$ symmetry). 
When the generated QFT is accidentally the same as the original one, these operations may be regarded as self-duality operations. 
The $SU(N)$ Yang--Mills theory with adjoint matter always has the self $\sfT$ duality associated with $\theta \to \theta+2\pi$ as we see in \eqref{eq:YM_anomaly}. 
Examples with the full $SL(2,\mathbb{Z})$ self-duality are the Cardy--Rabinovici model~\cite{Cardy:1981qy, Cardy:1981fd, Honda:2020txe, Hayashi:2022fkw} and $\mathcal{N}=4$ SYM theory~\cite{Vafa:1994tf, Donagi:1995cf, Dorey:1999sj, Polchinski:2000uf}. In Sec.~\ref{sec:N=1*SYM}, we shall discuss the $\mathcal{N}=1^*$ SYM theory in detail. 

\subsection{\texorpdfstring{$\sfS$}{S} and \texorpdfstring{$\sfT$}{T} operations on  't Hooft partition functions}

Let us study how these operations act on the 't~Hooft partition function~\eqref{eq:tHooftZ}. We define the $\sfS$ and $\sfT$ operations on $\calZtH[\Be,\Bm]$ by the temporal gauging of the $\sfS$- and $\sfT$-transformed partition functions, respectively:
\begin{align}
    &\sfS \calZtH[\Be,\Bm]\equiv \int \Diff a \exp\biggl(\frac{2\pi \im}{N}\int_{M_3} \Be\cup a\biggr) \sfS \calZ\biggl[\Bm+a\wedge \frac{\diff \tau}{L}\biggr], \\
    &\sfT \calZtH[\Be,\Bm]\equiv \int \Diff a \exp\biggl(\frac{2\pi \im}{N}\int_{M_3} \Be\cup a\biggr) \sfT \calZ\biggl[\Bm+a\wedge \frac{\diff \tau}{L}\biggr]. 
\end{align}
We can compute these path integrals explicitly to express the right-hand-side using $\calZtH$. 
The $\sfS$ transformation is given by 
\begin{align}
    \sfS \calZtH[\Be, \Bm]
    &=\int \Diff a\, \rme^{\frac{2\pi\im}{N}\int_{M_3}\Be a} \int \Diff b_m \Diff a'\, \rme^{\frac{2\pi \im}{N}\int_{M_3}(\Bm a'+b_m a)} \calZ\biggl[b_m+a'\wedge \frac{\diff \tau}{L}\biggr]\nonumber\\
    &=\int \Diff b_m \Diff a'\, N^{b_1(M_3)-1}\,\delta_{N}[\Be+b_m]\, \rme^{\frac{2\pi \im}{N}\int_{M_3}\Bm a'} \calZ\biggl[b_m+a'\wedge \frac{\diff \tau}{L}\biggr]\nonumber\\
    &=\calZtH[\Bm, -\Be]. 
    \label{eq:S_thooftZ}
\end{align}
Here, we decompose the dynamical $4$d $\mathbb{Z}_N$ gauge field as $b=b_m+a'\wedge \frac{\diff \tau}{L}$. 
The $\sfT$ transformation is given by
\begin{align}
    \sfT \calZtH[\Be, \Bm]
    &= \int \Diff a\, \rme^{\frac{2\pi \im}{N}\int_{M_3}\Be a}\rme^{\frac{2\pi \im}{N}\int_{M_3}\Bm a} \calZ\biggl[\Bm+a\wedge \frac{\diff \tau}{L}\biggr]\nonumber\\
    &=\calZtH[\Be+\Bm, \Bm]. 
    \label{eq:T_thooftZ}
\end{align}

Let us choose the generators $S,T\in SL(2,\mathbb{Z})$ to be 
\begin{equation}
    S=\begin{pmatrix}
        0&-1\\
        1&0
    \end{pmatrix}, \quad 
    T=\begin{pmatrix}
        1&-1\\
        0&1
    \end{pmatrix}, 
\end{equation}
so that $SL(2,\mathbb{Z})=\langle S,T\,|\, S^2=(ST^{-1})^3, S^4=1\rangle$ and $C\equiv S^2=(ST^{-1})^3$ corresponds to charge conjugation. 
If we write $\vec{B}=(\Be, \Bm)^t$ as a column vector, then the transformations~\eqref{eq:S_thooftZ} and \eqref{eq:T_thooftZ} can be expressed as 
\begin{align}
    \sfS \calZtH\Bigl[\vec{B}\Bigr] = \calZtH\Bigl[S^{-1} \vec{B}\Bigr],\quad 
    \sfT \calZtH\Bigl[\vec{B}\Bigr] = \calZtH\Bigl[T^{-1} \vec{B}\Bigr]. 
\end{align}
Thus, $\sfS$ and $\sfT$ transformations generate the $SL(2,\mathbb{Z})$ action on the space of $4$d QFTs with the $\mathbb{Z}_N^{[1]}$ symmetry. 
The explicit form of the action can be easily identified when we use the 't~Hooft partition function $\calZtH[\Be, \Bm]$, as we just need to perform an $SL(2,\mathbb{Z})$ transformation on the background field $\vec{B}=(\Be, \Bm)^t$. 
For example, we find 
\begin{equation}
    (\sfS^{p_1}\sfT^{q_1}\sfS^{p_2}\sfT^{q_2}\cdots)\calZtH[\vec{B}]
    =\calZtH[(S^{p_1}T^{q_2}S^{p_2}T^{q_2}\cdots)^{-1}\vec{B}].
\end{equation}
As we can express 
\begin{equation}
    (S^{p_1}T^{q_2}S^{p_2}T^{q_2}\cdots)=
    \begin{pmatrix}
        p & q \\
        r & s
    \end{pmatrix}\in SL(2,\mathbb{Z}), 
\end{equation}
with some $p,q,r,s \in \mathbb{Z}$ with $ps-qr=1$, the above relation becomes
\begin{equation}
    (\sfS^{p_1}\sfT^{q_1}\sfS^{p_2}\sfT^{q_2}\cdots)\calZtH[\Be, \Bm]
    =\calZtH[\,s \Be -q \Bm, -r \Be + p \Bm].
\end{equation}

Let us now consider two different gapped phases specified by order-$N$ subgroups $H_1$ and $H_2$, and assume that there is an isomorphism $H_1\iso H_2$ induced by an automorphism of $\widetilde{\mathbb{Z}}_N\times \mathbb{Z}_N$. 
Then, there should be an $SL(2,\mathbb{Z})$ operation that relates $H_1$ and $H_2$, and thus these different phases are connected in the web of $\sfS,\sfT$ operations. 

\subsection{Non-Abelian gapped phases in the \texorpdfstring{$\mathcal{N}=1^*$}{N=1*} SYM theory}
\label{sec:N=1*SYM}

${\cal N}=1^*$ SYM theory is defined as the mass deformation of the $4$d ${\cal N}=4$ SYM  theory.  It is one of the most interesting theoretical playgrounds for 4d theories as its vacuum structure is very rich~\cite{Donagi:1995cf, Dorey:1999sj, Polchinski:2000uf}. 
Our goal below is to express the 't~Hooft partition functions for the massive vacua of ${\cal N}=1^*$  theory. 

First, we provide a quick overview of ${\cal N}=1^*$ theory with $SU(N)$ gauge group starting with its ${\cal N}=4$  origin.
In the ${\cal N}=1$ notation, the field content of ${\cal N}=4$ SYM theory  is a vector multiplet $V$ and three chiral multiplets  $\Phi_i$ $(i=1, 2, 3)$ 
all in the adjoint representation of  $SU(N)$.  The action  of the theory  consists of gauge invariant kinetic terms of these fields plus a unique superpotential,
\begin{align}
 W= \frac{1}{g^2} \tr ( \Phi_1 [  \Phi_2,  \Phi_3]).
 \label{eq:superpotential}
\end{align}
The ${\cal N}=1^*$ theory is obtained from ${\cal N}=4$ by adding a mass term: 
 \begin{align}
 \Delta W=   \frac{m}{2 g^2} \left(  \tr \Phi_1^2   +  \tr \Phi_2^2  +\tr \Phi_3^2 \right).
\end{align}
In the $m \rightarrow \infty$ limit with arbitrarily small  coupling constant at the cut-off  $m$,  $ i \tau (m) \rightarrow  \infty$, with 
$ \Lambda^3 = m^3 \rme^{\im 2 \pi \tau(m)/N}$ fixed, the theory reduces to pure ${\cal N}=1$ theory, which is believed to be a confining gauge theory. At finite $m$, the theory has an extremely rich classical and quantum vacuum structure. 
Since we take the three masses non-zero,  the  moduli space of the ${\cal N}=4$ theory is completely lifted 
and the theory has  isolated vacua.  The classical  vacua are determined by the solutions of the $F$-term equations, $\partial W/\partial \Phi_i=0$, given by 
\begin{align}
[\Phi_i,  \Phi_j] =  -m \varepsilon_{ijk} \Phi_k .
\end{align}
Therefore, the supersymmetric classical vacua can be expressed by three $N  \times N$ matrices which obey the standard commutation relations for the $\mathfrak{su}(2)$ algebra. 
Up to gauge transformations, the classical vacua are described as 
\begin{equation}
    \Phi'_i\equiv\frac{1}{\im m}\Phi_i
    =\underbrace{J^{(d_1)}_i\oplus\cdots \oplus J^{(d_1)}_i}_{k_{d_1}}\oplus 
    \underbrace{J^{(d_2)}_i\oplus\cdots \oplus J^{(d_2)}_i}_{k_{d_2}}\oplus \cdots, 
\end{equation}
where $J^{(d)}_i$ are the generators of $d$-dimensional irreducible representation of $\mathfrak{su}(2)$, $k_d$ denotes its multiplicity, and $N=d_1k_{d_1}+d_2 k_{d_2}+\cdots$. 
Ignoring discrete factors momentarily,  the gauge structure at these vacua is reduced to $[\otimes_d U(k_d)]/U(1)$ at the classical level. 
If the classical vacuum contains different $\mathfrak{su}(2)$ representations, i.e. $N \neq d k_d$, then there will be unbroken $U(1)$'s in the infrared and it becomes a gapless Coulomb vacuum. 

Our primary interest in this paper is the 't Hooft partition functions of massive vacua, and they arise when the Higgs expectation values take the following form:
\begin{align}
    \Phi'_i=  \mathbf {1}_{N/d} \otimes  J_i^{(d)},  \qquad (i=1, 2, 3),
\end{align}
where $d$ is a divisor of $N$. As a result, the $SU(N)$ gauge group is Higgsed as 
\begin{equation}
    SU(N)\xrightarrow{\mathrm{Higgs}} \frac{SU(N/d)\times \mathbb{Z}_{N}}{\mathbb{Z}_{N/d}}, 
\end{equation}
where $SU(N/d)$ is the remaining continuous gauge group acting on the $\bm{1}_{N/d}$ component, $\mathbb{Z}_N$ in the numerator is the center subgroup of $SU(N)$, and they need to be divided by the common center $\mathbb{Z}_{N/d}$. 
As the classically massless contents describe the $\mathcal{N}=1$ $SU(N/d)$ SYM theory with emergent $\mathbb{Z}_{2(N/d)}$ discrete chiral symmetry, they are expected to be gapped by quantum fluctuations with $N/d$ distinct vacua. 
In addition to these confinement dynamics, the remnant discrete $\mathbb{Z}_N/\mathbb{Z}_{N/d}\simeq \mathbb{Z}_d$ gauge field describes the topological field theory for deconfined Wilson lines. 
As we have this dynamics for each divisor of $N$, the total number of the massive vacua is given by the divisor function,
\begin{align}
    \sigma (N)= \sum_{d|N} d  = \sum_{d|N} \left(\frac{N}{d} \right).
\end{align}
As pointed out in Ref.~\cite{Donagi:1995cf}, every gapped state in the Wilson--'t~Hooft classification can be realized as one of these vacua in the $\mathcal{N}=1^*$ SYM theory, and we will find their 't~Hooft partition functions and study how these vacua behave under $\sfS$ and $\sfT$ transformations.\footnote{A recent paper~\cite{Damia:2023ses} also studies the global properties of the $\mathcal{N}=1^*$ gapped vacua using the analogous $S$ and $T$ operations. } 
 
Depending on whether the prime factorization of $N$ is square free, the $SL(2,\mathbb{Z})$ orbit of the massive vacua shows different features. 
In the following, let us work on concrete examples, $N=6$ and $N=4$, as demonstrations.

\subsubsection*{Massive vacua of $\mathcal{N}=1^*$ $SU(6)$ SYM theory (the case with $N$ square-free)}
Let us consider $SU(6)$ gauge theory as a pedagogical example. 
Positive divisors of $N=6$ are $d=1,2,3$, and $6$, and we get the following table for the Higgs expectation values to have massive vacua:
\begin{align}
\begin{tabular}{ l | l | l | l} 
{\rm Higgs  vev} & $SU(6)  \xrightarrow{\mathrm{Higgs}}$  & $ (\mathbb Z_6^{[1]})_{\rm 4d} \xrightarrow{\mathrm{SSB}} $ &  $\calZtH[\Be,\Bm]$ \\ 
\hline
$J_i^{(d=6)}$  & $ \mathbb Z_6$  &  1 & $\delta_6[\Bm]$\\ 
$\bm{1}_2 \otimes   J_i^{(d=3)} $ & $ SU(2) \times \mathbb Z_3 $ &  $  (\mathbb Z_2^{[1]})_{\rm 4d}  $ & $\delta_6[2\Bm, 3\Be+k\Bm]$\\ 
$\bm{1}_3 \otimes   J_i^{(d=2)} $ & $ SU(3) \times \mathbb Z_2 $ &  $  (\mathbb Z_3^{[1]})_{\rm 4d}  $ & $\delta_6[3\Bm, 2\Be+k\Bm]$ \\  
$\bm{1}_6 \otimes   J_i^{(d=1)}=0 $ & $ SU(6)  $ &  $  (\mathbb Z_6^{[1]})_{\rm 4d}  $ & $\delta_6[\Be+k\Bm]$\\  
\end{tabular}
\end{align}
Let us discuss each of these vacuum structures to obtain the 't~Hooft partition functions and their relation under $SL(2,\mathbb{Z})$. 

Let us first discuss the Higgs phase, where $\Phi'_i=J_i^{(6)}$ and the gauge group is Higgsed to the center $\mathbb{Z}_6$. 
We note that this classical vacuum cannot be consistent with the nonzero magnetic flux $B_{4\rmd}\not=0$, and a vortex must be created to satisfy the boundary condition. Thus, the partition function with nonzero flux becomes exponentially small, 
\begin{equation}
    \calZ[B_{4\mathrm{d}}]=\delta_6[B_{4\rmd}],
\end{equation}
and the corresponding 't Hooft partition function is given by
\begin{equation}
    {\cal Z}_{\rm tH} [\Be, \Bm]= \delta_6[\Bm]. 
\end{equation}
Indeed, in this Higgs phase, the electrically charged particles associated with the triplet of adjoint scalars condense, and we naturally expect the confinement of magnetic charges. 

In the totally confined phase, the Higgs expectation values are zero, $\Phi'_i=\bm{1}\otimes J^{(1)}_i=0$, and the $\mathbb{Z}_6^{[1]}$ symmetry is unbroken. 
At the classical level, gapless gluons and gluinos are associated with the $SU(6)$ gauge group with $\mathcal{N}=1$ supersymmetry, and quantum fluctuations should generate confinement. 
In the $\mathcal{N}=1^*$ theory, the counterpart of discrete chiral symmetry of $\mathcal{N}=1$ SYM does not exist as the potentially anomalous $U(1)$ symmetry is broken at the classical level by the superpotential~\eqref{eq:superpotential}, but the low-energy effective theory for the confinement phase acquires an emergent $\mathbb{Z}_{2N}$ axial symmetry. 
As computed in \eqref{eq:chiralbrokenvacua}, the  't Hooft partition functions for these gapped chiral-broken vacua are given by
 \begin{equation}
    {\cal Z}_{\rm tH} [\Be, \Bm]= \delta_6[\Be + k \Bm],
\end{equation}
where $k=1,\ldots, N$ specifies the phase of the gluino condensate and they are cyclically permuted by the $\sfT$ operation. 
Another important remark is that the $k=0$ confining state is dual to the Higgs state by the $\sfS$ operation as expected from electromagnetic duality. 

 \begin{figure}[t]
\vspace{-1.0cm}
\centering
\includegraphics[width = 0.85\textwidth]{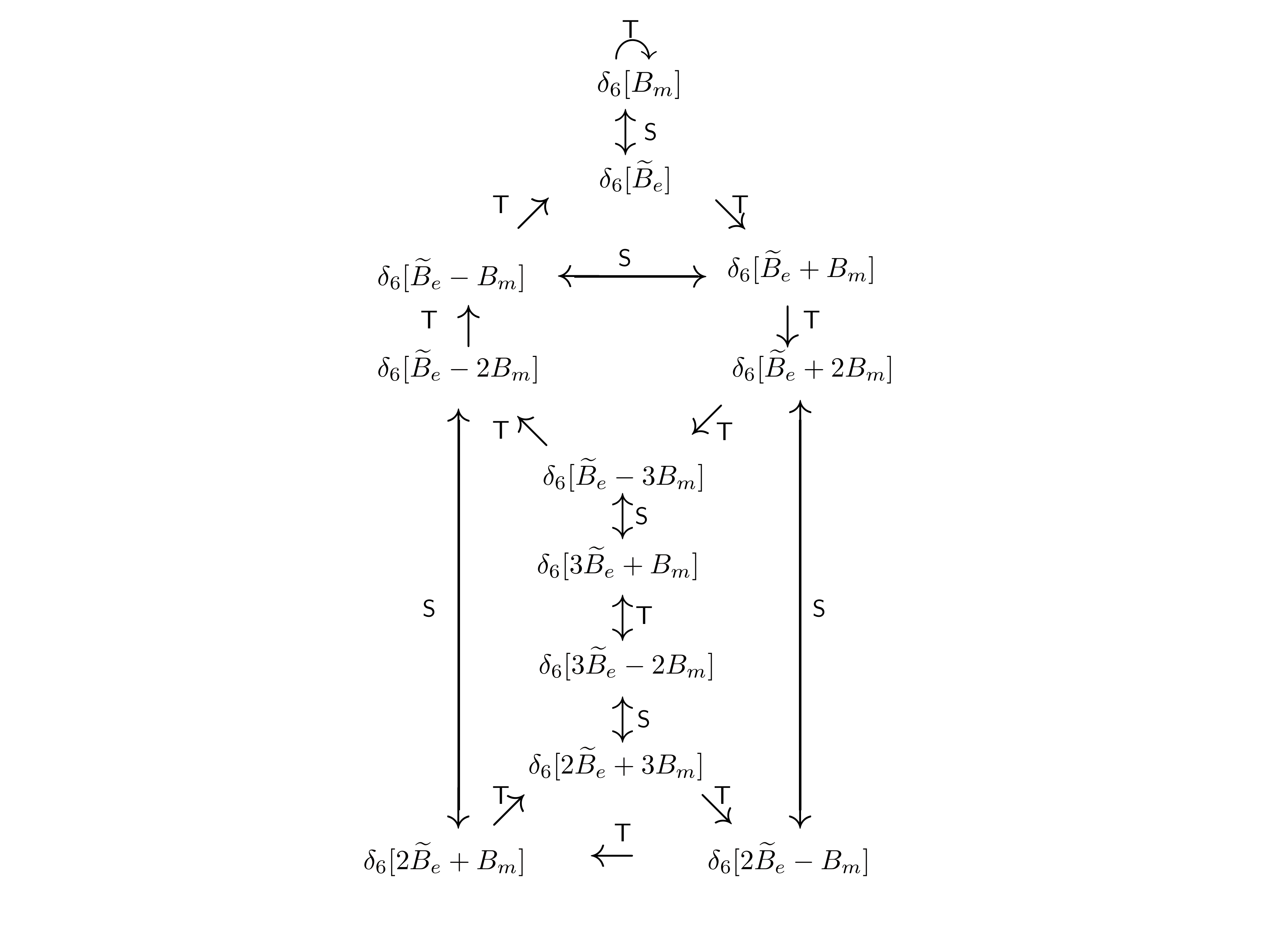}
\vspace{-0.7cm}
\caption{ 't Hooft partition functions for  massive vacua of $SU(6)$  $\mathcal{N}=1^{*}$ SYM theory,  and their $SL(2,{\mathbb Z})$  transformations.  There are $\sigma(6)=1+2+3+6=12$ supersymmetric gapped vacua and these phases are in 1:1 correspondence with the symmetry breaking pattern 
$  (\mathbb{Z}_N^{[1]})_{4\mathrm{d}}\rightarrow (\mathbb{Z}_n^{[1]})_{4\mathrm{d}} $ 
  enriched with the $\mathbb{Z}_n^{[1]}$ level-$k$ SPT state. 
Since $N=6$ is a square-free integer, all massive phases are connected by $\mathsf{S}$ and $\mathsf{T}$ transformations.
}
\label{fig:SU(6)PF}
\end{figure}

The analyses of the remaining two cases are similar, so let us focus on the case of $\Phi'_i=\bm{1}_2\otimes J^{(3)}_i$, where the gauge group becomes
\begin{equation}
    SU(6)\xrightarrow{\mathrm{Higgs}} SU(2)\times \mathbb{Z}_3. 
    \label{eq:higgsing6to2}
\end{equation}
The Higgs expectation value $\Phi'_i=\bm{1}_2\otimes J^{(3)}_i$ is compatible with the magnetic fluxes $\int_{\Sigma}\frac{2\pi}{6}B_{4\rmd}\in \pi\mathbb{Z}$, so the partition function should contain the factor $\delta_6[2 B_{4\rmd}]$. 
The possible 't~Hooft partition functions for gapped phases are then given by 
\begin{equation}
    \delta_6[2\Bm, 3\Be+k\Bm],
    \label{eq:Zth6to2}
\end{equation}
with $k\sim k+2$. We note that these two states are related by the $\sfT$ operation since 
\begin{equation}
    \sfT\colon \delta_6[2\Bm, 3\Be]\longleftrightarrow \delta_6[2\Bm, 3\Be+3\Bm]=\delta_6[2\Bm, 3\Be+\Bm].
\end{equation}
Let us discuss why this is the case. To proceed, we need to study the confinement dynamics of the effective $\mathcal{N}=1$ $SU(2)$ SYM theory, and we should note that the $\theta$ angle for the $SU(2)$ gauge group becomes $\theta_{\mathrm{eff}}=3\theta$ in the Higgsing \eqref{eq:higgsing6to2}. 
The $\sfT$ operation, $\theta\to \theta+2\pi$, acts as $\theta_{\mathrm{eff}}\to \theta_{\mathrm{eff}}+6\pi =\theta_{\mathrm{eff}}+2\pi \bmod 4\pi$ and the chiral broken vacua for the $SU(2)$ theory are actually related by the $\sfT$ transformation.

The complete list of the 't~Hooft partition functions for gapped phases are listed in Fig.~\ref{fig:SU(6)PF}, and we also show how the $SL(2,\mathbb{Z})$ operations relate those phases.\footnote{This figure has the identical structure with the $S,T$-duality orbit for $\mathcal{N}=4$ SYM of the gauge Lie algebra $\mathfrak{g}=\mathfrak{su}(6)$ in Ref.~\cite{Aharony:2013hda}, and some readers may be confused about the difference between them. The key difference is that Ref.~\cite{Aharony:2013hda} discusses the global structure of the gauge group itself, which is about the kinematics not the dynamics, while we here discuss the vacuum structure of the $SU(6)$ gauge theory.  In particular, in Ref.~\cite{Aharony:2013hda}, one constructs genuine line operators in $(SU(N)/\mathbb Z_n)_k$ theory, while in our case, we are depicting the dyonic charges that are screened in the $SU(N)$ theory.  } 
All the gapped states are exchanged by $\sfS$ and $\sfT$ operations. 
This is a general fact when $N$ is a square-free integer, and we demonstrate it here for $N=6$ as an example. 
In this figure, we use the fact that the 't~Hooft partition function for this case can be written using a single delta functional. 
For example, \eqref{eq:Zth6to2} contains two different delta functions, but it can be equivalently rewritten as  
\begin{equation}
    \delta_6[2\Bm, 3\Be+k\Bm]=\delta_6[3\Be+k\Bm]
\end{equation}
for $k=1,2$ (Since $k\sim k+2$, we can always choose this way). 
This is related to the fact that all the order $N$ subgroups of $\widetilde{\mathbb{Z}}_N\times \mathbb{Z}_N$ is isomorphic to $\mathbb{Z}_N$ when $N$ is square free.\footnote{Let us prove this fact using the 't~Hooft partition function (in a physically intuitive way). We start from the 't~Hooft partition function, $\delta_N[n\Bm, \frac{N}{n}\Be+k\Bm]$, and we assume that $\gcd(n, N/n)=1$, which is always the case if $N$ is square free. The $\sfT$ transformation shifts $k\to k+(N/n)$, and this surveys all possible $k\sim k+n$ due to $\gcd(n,N/n)=1$. In particular, we can reach the vacuum, $\delta_N[n \Bm, \frac{N}{n}\Be+\Bm]=\delta_N[\frac{N}{n}\Be+\Bm]$, and its $\sfS$ transformation gives one of the totally confined phase, $\delta_N[\Be-\frac{N}{n}\Bm]$. } 

\subsubsection*{Massive vacua of $\mathcal{N}=1^*$ $SU(4)$ SYM theory (the case with $N$ not square-free)}

When $N$ contains squares in its prime factorization, the $SL(2,\mathbb{Z})$ structure of massive vacua has disconnected components. Let us discuss $N=4$ as the simplest example.  
The massive vacua are listed in the following table:
\begin{align}
\begin{tabular}{ l | l | l | l} 
{\rm Higgs  vev} & $SU(4)  \xrightarrow{\mathrm{Higgs}}$  & $ (\mathbb Z_4^{[1]})_{\rm 4d} \xrightarrow{\mathrm{SSB}} $ &  $\calZtH[\Be,\Bm]$ \\ 
\hline
$J_i^{(d=4)}$  & $ \mathbb Z_4$  &  1 & $\delta_4[\Bm]$\\ 
$\bm{1}_2 \otimes   J_i^{(d=2)} $ & $ [SU(2) \times \mathbb Z_4]/\mathbb{Z}_2 $ &  $  (\mathbb Z_2^{[1]})_{\rm 4d}  $ & $\delta_4[2\Bm, 2\Be+k\Bm]$ \\  
$\bm{1}_4 \otimes   J_i^{(d=1)}=0 $ & $ SU(4)  $ &  $  (\mathbb Z_4^{[1]})_{\rm 4d}  $ & $\delta_4[\Be+k\Bm]$\\  
\end{tabular}
\end{align}
The analysis of the Higgs and totally confining phases is completely the same as we have done for $N=6$, so let us here focus on the case $\Phi'_i=\bm{1}_2\otimes J_i^{(2)}$. 

\begin{figure}[t]
\vspace{-1.0cm}
\centering
\includegraphics[width = 0.85\textwidth]{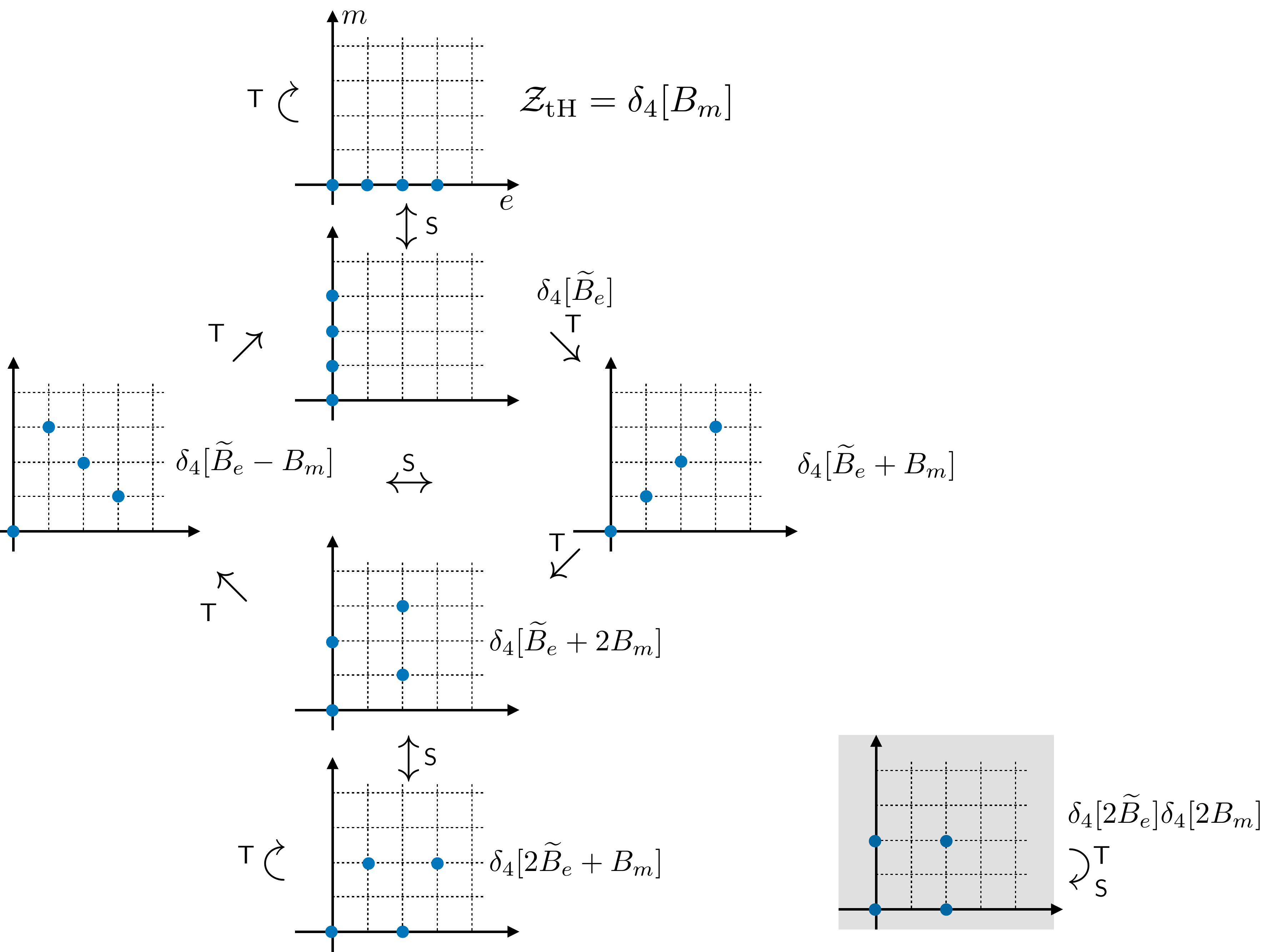}
\vspace{-0.1cm}
\caption{ Phases and 't Hooft partition functions for the ${\cal N}=1^{*}$ SYM theory  with $SU(4)$ gauge group. 
The marked points on the lattice indicate the electromagnetic charges of deconfined dyonic lines in each phase.  
As $N=4$ is not square free, the web of $\mathsf{S}$ and $\mathsf{T}$ transformations is disconnected. }
\label{fig:SU(4)PF}
\end{figure}

The Higgs expectation value $\Phi'_i=\bm{1}_2\otimes J_i^{(2)}$ causes the Higgsing of the gauge group, 
\begin{equation}
    SU(4)\xrightarrow{\mathrm{Higgs}}\frac{SU(2)\times \mathbb{Z}_4}{\mathbb{Z}_2}, 
\end{equation}
and it breaks 1-form symmetry as 
$(\mathbb Z_4^{[1]})_{\rm 4d} \rightarrow (\mathbb Z_2^{[1]})_{\rm 4d} $. 
The $SU(2)$ gauge group is confined in the infrared and it exhibits two vacua due to the spontaneous breaking of emergent chiral symmetry. These two phases are distinguished by the SPT actions of the unbroken $(\mathbb{Z}_2^{[1]})_{4\rmd}$ symmetry, and thus their $4$d partition functions are given by 
\begin{equation}
    \calZ[B_{4\rmd}]=\delta_4[2B_{4\rmd}]\exp\biggl(\frac{2\pi \im k}{2}\int \frac{1}{2}P_2(B_{4\rmd}/2)\biggr),
\end{equation}
with $k\sim k+2$. The corresponding 't~Hooft partition functions are given by
\begin{equation}
    \calZtH[\Be, \Bm]=\delta_4[2\Bm, 2\Be+k\Bm]. 
\end{equation}
Importantly, these chiral broken vacua are \textit{not} related by the $\sfT$ transformation, $\Be\to \Be+\Bm$, and each chiral-broken vacuum is invariant under $\sfT$. 
To understand their $\sfT$ invariance from the microscopic viewpoint, we should notice that the effective $\theta$ angle of the $SU(2)$ theory is given by $\theta_{\mathrm{eff}}=2\theta$ using the $\theta$ angle of the $SU(4)$ theory. 
Under the $2\pi$ shift of $\theta$, the effective $\theta$ angle is shifted as $\theta_{\mathrm{eff}}\to \theta_{\mathrm{eff}}+4\pi$, and each chiral-broken vacua of $\mathcal{N}=1$ $SU(2)$ SYM theory is invariant under this operation.

In Fig.~\ref{fig:SU(4)PF}, we give the complete list of the 't~Hooft partition functions for gapped vacua and their relations under the $\sfS$ and $\sfT$ operations. 
We also show the set of deconfined lines of each gapped vacua. 
Unlike the $SU(6)$ case, there are disconnected components in the duality web of $\sfS$ and $\sfT$ operations for $SU(4)$ theory, and in particular, $\calZtH[\Be,\Bm]=\delta_4[2\Bm,2\Be]$ is invariant under both $\sfS$ and $\sfT$ operations. 
The presence of disconnected components is a general feature for cases with $N$ not square free, and $SU(4)$ is an illustrative example.

\section{Summary and discussion}
\label{sec:discussion}

In this paper, we introduce the temporal gauging for $4$d QFTs with $\mathbb{Z}_N^{[1]}$ symmetry and define the 't~Hooft partition function. 
This operation does not respect the $4$d Lorentz invariance, but it introduces the spatial $\widetilde{\mathbb{Z}}_N^{[1]}\times \mathbb{Z}_N^{[1]}$ symmetry, and thus the spatial Wilson and 't~Hooft lines become genuine line operators (while all the temporal line operators are no longer genuine). 
This allows us to justify the classification of gapped phases using the Wilson--'t~Hooft criterion, and we establish its $1$-to-$1$ correspondence with the spontaneous breakdown of $4$d $\mathbb{Z}_N^{[1]}$ symmetry enriched with the SPT phase of the unbroken $1$-form symmetry. 

In other words, our argument justifies the use of dyonic line operators to classify the SPT phases for the vacua of $4$d gauge theories. 
This may be reminiscent of the use of the string order parameter~\cite{PhysRevB.40.4709} to characterize the Haldane gap or AKLT state~\cite{Haldane:1983ru, Affleck:1987vf}. 
The Kennedy--Tasaki (KT) transformation~\cite{kennedy1992hidden, Kennedy:1992ifl, oshikawa1992hidden} maps this nonlocal string order parameter to a local correlation function, and the SPT nature of the AKLT state can be understood as the spontaneous breaking of hidden (or dual) $\mathbb{Z}_2\times \mathbb{Z}_2$ symmetry (see Ref.~\cite{Li:2023mmw} for its field-theoretic description from a modern viewpoint). 
In this context, we may say that the temporal gauging of $4$d $\mathbb{Z}_N^{[1]}$ symmetry gives a suitable KT transformation for $4$d gauge theories to classify their gapped phases solely by spontaneous symmetry breaking. 
Honestly, it is quite astonishing that 't~Hooft had already introduced all the essential ingredients in Refs.~\cite{tHooft:1977nqb, tHooft:1979rtg, tHooft:1981bkw} before any of these developments. 

In this paper, we compute the 't~Hooft partition function for the simplest $4$d topological states and make the connection with the Wilson--'t~Hooft criterion. 
As the general $4$d topological states are known to be described by the $2$-group gauge theories~\cite{Lan:2018vjb, Johnson-Freyd:2020usu,  Thorngren:2020aph}, it would be an interesting future study to compute the 't~Hooft partition functions of these topological states, which would give us better understanding of $4$d gapped phases.  

\acknowledgments
 
The work of Y.~T. was supported by Japan Society for the Promotion of Science (JSPS) KAKENHI Grant numbers, 22H01218, and by Center for Gravitational Physics and Quantum Information (CGPQI) at Yukawa Institute for Theoretical Physics.
The work of M.~\"{U}. was supported by U.S. Department of Energy, Office of Science, Office of Nuclear Physics under Award Number DE-FG02-03ER41260.

\appendix

\section{Duality equation on \texorpdfstring{$T^4$}{T4} and classification of \texorpdfstring{$4$d}{4d} gapped phases}
\label{sec:tHooftoriginal}

In the main text, we have taken the point of view that the mutual locality condition \eqref{eq:mutuallocality} is nothing but the statement that the spontaneous breakdown of 1-form symmetry be consistent with 4d Lorentz invariance. Here, we review 't~Hooft's original derivation of this condition \cite{tHooft:1979rtg}, which is quite insightful and also interesting for its elementary character. 

We take the Euclidean spacetime manifold to be a flat four-torus with $L_{\mu = 1,\ldots,4}$ the circumferences of the various circles in $T^4 = (S^1)^4$. With this topology, the background gauge fields $A$, $\Bm$, $\Be$ are associated with triplets of mod $N$ integers $\Vec{n} = (n_1,n_2,n_3)\equiv (n_{14},n_{24},n_{34})$, $\vec{m} = (m_1,m_2,m_3)\equiv (n_{23},n_{31},n_{12})$, $\Vec{e}= (e_1, e_2, e_3)\equiv (e_{23},e_{31},e_{12})$ via
\begin{align}
    B_{4\rmd}&=\Bm+A\wedge \frac{\diff \tau}{L_4}=\sum_{i<j} n_{ij} \frac{\diff x_i \wedge \diff x_j}{L_i L_j}+\sum_{i} n_{i4} \frac{\diff x_i\wedge \diff \tau}{L_iL_4},\\
    \Be &= \sum_{i<j}  e_{ij} \frac{\diff x_i \wedge \diff x_j}{L_i L_j}.
\end{align}
Accordingly, we shall write the ordinary and 't Hooft partition functions as
\begin{align}
    \calZ[B_{4\rmd}] \equiv \calZ[\Vec{n},\Vec{m}], \quad \calZtH[\Be,\Bm] \equiv \calZtH[\Vec{e}, \Vec{m}],
\end{align}
and the relation between them as\footnote{When we use the definition~\eqref{eq:tHooftZ}, the canonical normalization is given by $\frac{1}{|H^0(M_3;\mathbb{Z}_N)|}=\frac{1}{N}$ instead of $\frac{1}{N^3}$, and the inverse operation $\int \Diff \Be$ has the normalization factor $\frac{|H^0(M_3;\mathbb{Z}_N)|}{|H^1(M_3;\mathbb{Z}_N)|}=\frac{1}{N^{\beta_1(M_3)-1}}=\frac{1}{N^2}$. Here, we follow the original normalization by 't~Hooft that assigns $1/N^3$ for the 't~Hooft partition function. 
The following argument basically works in both normalizations, but we would like here to note that the normalization of \eqref{eq:tHooftZ_GappedPhases} becomes $1$ instead of $N^{\beta_1(M_3)-1}$ in this convention. }
\begin{align}
    \calZtH[\Vec{e},\Vec{m}] = \frac{1}{N^3}\sum_{\Vec{n}} \exp \biggl( \frac{2\pi \im }{N}\Vec{e} \cdot \Vec{n} \biggr) \,\calZ[\Vec{n},\Vec{m}].
\end{align}

4d Lorentz invariance provides a important constraint on the 't Hooft partition function. In particular, let us consider the Euclidean Lorentz transformation
\begin{equation}
    \Lambda = 
    \begin{pmatrix}
    0 & 1 & 0 & 0 \\
    -1 & 0 & 0 & 0 \\
    0 & 0 & 0 & -1 \\
    0 & 0 & 1 & 0 
    \end{pmatrix},
\end{equation}
then the magnetic flux is transformed as $(n'_{ij})=\Lambda^t (n_{ij})\Lambda$. 
This has the effect of interchanging the pairs $(n_1,n_2)$ and $(m_1,m_2)$ as we have 
\begin{equation}
    \begin{pmatrix}
        -m'_2& n'_1\\
        m'_1 & n'_2
    \end{pmatrix}
    =
    \begin{pmatrix}
        0& -1\\
        1 & 0
    \end{pmatrix}
    \begin{pmatrix}
        -m_2& n_1\\
        m_1 & n_2
    \end{pmatrix}
    \begin{pmatrix}
        0& -1\\
        1 & 0
    \end{pmatrix},
\end{equation}
while keeping $n_3,m_3$ fixed. Using the notation $\Tilde{q} \equiv (q_1,q_2)$ for a three-vector $\vec{q} = (q_1,q_2,q_3)$, the covariance of the ordinary partition function under the above Lorentz transformation reads (we also have $L_1 \leftrightarrow L_2, L_3 \leftrightarrow L_4$, but this is suppressed in our notation)
\begin{align}
    \calZ [\Tilde{n},n_3;\Tilde{m},m_3] = \calZ [\Tilde{m},n_3; \Tilde{n},m_3].
\end{align}
After Fourier transformation, we then obtain
\begin{align}
    \calZtH [\tilde{e},e_3; \Tilde{m},m_3]
    = \frac{1}{N^2} \sum_{\Tilde{e}', \Tilde{m}'} \exp \biggl( \frac{2\pi \im}{N}( \tilde{e} \cdot \tilde{m}' - \tilde{m} \cdot \tilde{e}') \biggr) \, \calZtH[\Tilde{e}',e_3;\tilde{m}',m_3],
    \label{eq:tHooft_Duality}
\end{align}
which is what 't Hooft calls the ``duality relation.''

To obtain the constraints on the gapped vacua, 't Hooft makes a technical assumption: If the vacuum is gapped, then the ratio $\calZtH[\Vec{e},\Vec{m}]/\calZtH[\Vec{0},\Vec{0}]$ should approach either $0$ or $1$ as $L_{\mu = 1,\ldots,4} \to \infty$. 
Although this seems to be a nontrivial assumption, we have checked in \eqref{eq:tHooftZ_GappedPhases} that it is actually valid for the $\mathbb{Z}_{N/n}$ topological states with a $\mathbb{Z}_n$ SPT phase. 
It would be nice if we could prove/disprove it for general $4$d topological states with $\mathbb{Z}_N^{[1]}$. 

Under the above assumption, a gapped phase is then characterized by the set of all fluxes that are `light,' i.e., the set of fluxes $(\Vec{e},\vec{m})$ with $\calZtH[\Vec{e},\Vec{m}]/\calZtH[\Vec{0},\Vec{0}] \to 1$. 
The possible sets of light fluxes are strongly constrained by the duality relation \eqref{eq:tHooft_Duality}. 
As shown in Ref.~\cite{tHooft:1979rtg}, any two light fluxes $(\Vec{e},\Vec{m}),(\Vec{e}\,',\Vec{m}')$ with $e_3 = e_3{}', m_3,=m_3{}'$ must satisfy
\begin{align}
    \tilde{e} \cdot \tilde{m}'- \tilde{m} \cdot \tilde{e}' = 0 \mod N,
\end{align}
and there are either exactly $N^2$ or $0$ light fluxes out of the possible $N^4$ fluxes with given $e_3,m_3$. 

\noindent\textbf{[Proof]} Let us fix the normalization $\calZtH[\Vec{0},\Vec{0}] = 1$.
Suppose $(\tilde{e},e_3; \Tilde{m},m_3)$ is a light flux. 
Let us first establish that $(\tilde{0},e_3; \Tilde{0},m_3)$ is also light. To see this, we note that
\begin{align}
    1&= \calZtH [\tilde{e},e_3; \Tilde{m},m_3] 
    = \frac{1}{N^2} \sum_{\Tilde{e}', \Tilde{m}'} \exp \biggl( \frac{2\pi \im}{N}( \tilde{e} \cdot \tilde{m}' - \tilde{m} \cdot \tilde{e}') \biggr) \, \calZtH[\Tilde{e}',e_3;\tilde{m}',m_3] \nonumber\\
    &\leq \frac{1}{N^2} \sum_{\Tilde{e}', \Tilde{m}'} \calZtH[\Tilde{e}',e_3;\tilde{m}',m_3]
    = \calZtH[\Tilde{0},e_3;\tilde{0},m_3] \leq 1,
    \label{eq:squeeze}
\end{align}
where the second step uses the duality relation, the third step uses the positivity of $\calZtH$, and the fourth step uses the duality relation again. So indeed $\calZtH[\Tilde{0},e_3;\tilde{0},m_3] = 1$. 

Replacing the final inequality with an equality in \eqref{eq:squeeze}, we find 
\begin{align}
    \frac{1}{N^2} \sum_{\Tilde{e}', \Tilde{m}'} \calZtH[\Tilde{e}',e_3;\tilde{m}',m_3]
    = \calZtH[\Tilde{0},e_3;\tilde{0},m_3] = 1.
\end{align}
Clearly, this can hold only if precisely $N^2$ of the $N^4$ terms on the left-hand side are equal to $1$, with all the remaining terms equal to $0$. In other words, we have established that precisely $N^2$ among the $N^4$ fluxes $(\Tilde{e}',e_3;\tilde{m}',m_3)$ are light. 

Let us now look at the duality relation again,
\begin{align*}
    1 = \calZtH [\tilde{e},e_3; \Tilde{m},m_3]
    = \frac{1}{N^2} \sum_{\Tilde{e}', \Tilde{m}'} \exp \biggl( \frac{2\pi \im}{N}( \tilde{e} \cdot \tilde{m}' - \tilde{m} \cdot \tilde{e}') \biggr) \, \calZtH[\Tilde{e}',e_3;\tilde{m}',m_3].
\end{align*}
As exactly $N^2$ of the $\calZtH [\tilde{e}',e_3;\tilde{m}',m_3]$ equal 1 while all the others vanish, this equality can be true iff $\exp \bigl(\frac{2\pi \im}{N}( \tilde{e} \cdot \tilde{m}' - \tilde{m} \cdot \tilde{e}') \bigr) = 1$ for each light flux $(\tilde{e}',e_3;\tilde{m}',m_3)$. 
$\square$ 


\bibliographystyle{utphys}
\bibliography{./QFT.bib,./refs.bib}

\providecommand{\href}[2]{#2}\begingroup\raggedright\begin{thebibliography}{10}

\bibitem{Wilson:1974sk}
K.~G. Wilson, ``{Confinement of Quarks},''
\href{http://dx.doi.org/10.1103/PhysRevD.10.2445}{{\em Phys. Rev.} {\bfseries
  D10} (1974) 2445--2459}.

\bibitem{tHooft:1977nqb}
G.~'t~Hooft, ``On the phase transition towards permanent quark confinement,''
  \href{http://dx.doi.org/10.1016/0550-3213(78)90153-0}{{\em Nucl.Phys.B}
  {\bfseries 138} (1978) 1--25}.

\bibitem{tHooft:1979rtg}
G.~'t~Hooft, ``{A Property of Electric and Magnetic Flux in Nonabelian Gauge
  Theories},''
\href{http://dx.doi.org/10.1016/0550-3213(79)90595-9}{{\em Nucl. Phys.}
  {\bfseries B153} (1979) 141--160}.

\bibitem{tHooft:1981bkw}
G.~'t~Hooft, ``{Topology of the Gauge Condition and New Confinement Phases in
  Nonabelian Gauge Theories},''
\href{http://dx.doi.org/10.1016/0550-3213(81)90442-9}{{\em Nucl. Phys.}
  {\bfseries B190} (1981) 455--478}.

\bibitem{Aharony:2013hda}
O.~Aharony, N.~Seiberg, and Y.~Tachikawa, ``{Reading between the lines of
  four-dimensional gauge theories},''
  \href{http://dx.doi.org/10.1007/JHEP08(2013)115}{{\em JHEP} {\bfseries 08}
  (2013) 115},
\href{http://arxiv.org/abs/1305.0318}{{\ttfamily arXiv:1305.0318 [hep-th]}}.

\bibitem{Kapustin:2014gua}
A.~Kapustin and N.~Seiberg, ``{Coupling a QFT to a TQFT and Duality},''
  \href{http://dx.doi.org/10.1007/JHEP04(2014)001}{{\em JHEP} {\bfseries 04}
  (2014) 001},
\href{http://arxiv.org/abs/1401.0740}{{\ttfamily arXiv:1401.0740 [hep-th]}}.

\bibitem{Gaiotto:2014kfa}
D.~Gaiotto, A.~Kapustin, N.~Seiberg, and B.~Willett, ``{Generalized Global
  Symmetries},'' \href{http://dx.doi.org/10.1007/JHEP02(2015)172}{{\em JHEP}
  {\bfseries 02} (2015) 172},
\href{http://arxiv.org/abs/1412.5148}{{\ttfamily arXiv:1412.5148 [hep-th]}}.

\bibitem{Gaiotto:2017yup}
D.~Gaiotto, A.~Kapustin, Z.~Komargodski, and N.~Seiberg, ``{Theta, Time
  Reversal, and Temperature},''
  \href{http://dx.doi.org/10.1007/JHEP05(2017)091}{{\em JHEP} {\bfseries 05}
  (2017) 091},
\href{http://arxiv.org/abs/1703.00501}{{\ttfamily arXiv:1703.00501 [hep-th]}}.

\bibitem{Shimizu:2017asf}
H.~Shimizu and K.~Yonekura, ``{Anomaly constraints on deconfinement and chiral
  phase transition},'' \href{http://dx.doi.org/10.1103/PhysRevD.97.105011}{{\em
  Phys. Rev.} {\bfseries D97} no.~10, (2018) 105011},
\href{http://arxiv.org/abs/1706.06104}{{\ttfamily arXiv:1706.06104 [hep-th]}}.

\bibitem{Tanizaki:2017qhf}
Y.~Tanizaki, T.~Misumi, and N.~Sakai, ``{Circle compactification and 't Hooft
  anomaly},'' \href{http://dx.doi.org/10.1007/JHEP12(2017)056}{{\em JHEP}
  {\bfseries 12} (2017) 056},
\href{http://arxiv.org/abs/1710.08923}{{\ttfamily arXiv:1710.08923 [hep-th]}}.

\bibitem{Tanizaki:2017mtm}
Y.~Tanizaki, Y.~Kikuchi, T.~Misumi, and N.~Sakai, ``{Anomaly matching for phase
  diagram of massless $\mathbb{Z}_N$-QCD},''
  \href{http://dx.doi.org/10.1103/PhysRevD.97.054012}{{\em Phys. Rev.}
  {\bfseries D97} (2018) 054012},
\href{http://arxiv.org/abs/1711.10487}{{\ttfamily arXiv:1711.10487 [hep-th]}}.

\bibitem{Gukov:2013zka}
S.~Gukov and A.~Kapustin, ``{Topological Quantum Field Theory, Nonlocal
  Operators, and Gapped Phases of Gauge Theories},''
  \href{http://arxiv.org/abs/1307.4793}{{\ttfamily arXiv:1307.4793 [hep-th]}}.

\bibitem{Kapustin:2013uxa}
A.~Kapustin and R.~Thorngren, ``{Higher symmetry and gapped phases of gauge
  theories},'' \href{http://arxiv.org/abs/1309.4721}{{\ttfamily arXiv:1309.4721
  [hep-th]}}.

\bibitem{Kapustin:2013qsa}
A.~Kapustin and R.~Thorngren, ``{Topological Field Theory on a Lattice,
  Discrete Theta-Angles and Confinement},''
  \href{http://dx.doi.org/10.4310/ATMP.2014.v18.n5.a4}{{\em Adv. Theor. Math.
  Phys.} {\bfseries 18} no.~5, (2014) 1233--1247},
  \href{http://arxiv.org/abs/1308.2926}{{\ttfamily arXiv:1308.2926 [hep-th]}}.

\bibitem{Lan:2018vjb}
T.~Lan, L.~Kong, and X.-G. Wen, ``{Classification of (3+1)D Bosonic Topological
  Orders: The Case When Pointlike Excitations Are All Bosons},''
  \href{http://dx.doi.org/10.1103/PhysRevX.8.021074}{{\em Phys. Rev. X}
  {\bfseries 8} no.~2, (2018) 021074}.

\bibitem{Johnson-Freyd:2020usu}
T.~Johnson-Freyd, ``{On the Classification of Topological Orders},''
  \href{http://dx.doi.org/10.1007/s00220-022-04380-3}{{\em Commun. Math. Phys.}
  {\bfseries 393} no.~2, (2022) 989--1033},
  \href{http://arxiv.org/abs/2003.06663}{{\ttfamily arXiv:2003.06663
  [math.CT]}}.

\bibitem{Thorngren:2020aph}
R.~Thorngren, ``{Topological quantum field theory, symmetry breaking, and
  finite gauge theory in 3+1D},''
  \href{http://dx.doi.org/10.1103/PhysRevB.101.245160}{{\em Phys. Rev. B}
  {\bfseries 101} no.~24, (2020) 245160},
  \href{http://arxiv.org/abs/2001.11938}{{\ttfamily arXiv:2001.11938
  [cond-mat.str-el]}}.

\bibitem{Tanizaki:2017bam}
Y.~Tanizaki and Y.~Kikuchi, ``{Vacuum structure of bifundamental gauge theories
  at finite topological angles},''
  \href{http://dx.doi.org/10.1007/JHEP06(2017)102}{{\em JHEP} {\bfseries 06}
  (2017) 102},
\href{http://arxiv.org/abs/1705.01949}{{\ttfamily arXiv:1705.01949 [hep-th]}}.

\bibitem{Karasik:2019bxn}
A.~Karasik and Z.~Komargodski, ``The bi-fundamental gauge theory in 3+1
  dimensions: The vacuum structure and a cascade,''
  \href{http://dx.doi.org/10.1007/JHEP05(2019)144}{{\em JHEP} {\bfseries 05}
  (2019) 144}, \href{http://arxiv.org/abs/1904.09551}{{\ttfamily
  arXiv:1904.09551 [hep-th]}}.

\bibitem{Cordova:2019uob}
C.~Cordova, D.~S. Freed, H.~T. Lam, and N.~Seiberg, ``{Anomalies in the Space
  of Coupling Constants and Their Dynamical Applications II},''
  \href{http://dx.doi.org/10.21468/SciPostPhys.8.1.002}{{\em SciPost Phys.}
  {\bfseries 8} no.~1, (2020) 002},
  \href{http://arxiv.org/abs/1905.13361}{{\ttfamily arXiv:1905.13361
  [hep-th]}}.

\bibitem{Witten:1979ey}
E.~Witten, ``{Dyons of Charge $e\theta/2\pi$},''
\href{http://dx.doi.org/10.1016/0370-2693(79)90838-4}{{\em Phys. Lett.}
  {\bfseries B86} (1979) 283--287}.

\bibitem{Sharpe:2015mja}
E.~Sharpe, ``{Notes on generalized global symmetries in QFT},''
  \href{http://dx.doi.org/10.1002/prop.201500048}{{\em Fortsch. Phys.}
  {\bfseries 63} (2015) 659--682},
\href{http://arxiv.org/abs/1508.04770}{{\ttfamily arXiv:1508.04770 [hep-th]}}.

\bibitem{Tachikawa:2017gyf}
Y.~Tachikawa, ``{On gauging finite subgroups},''
  \href{http://dx.doi.org/10.21468/SciPostPhys.8.1.015}{{\em SciPost Phys.}
  {\bfseries 8} (2020) 015},
\href{http://arxiv.org/abs/1712.09542}{{\ttfamily arXiv:1712.09542 [hep-th]}}.

\bibitem{Cordova:2018cvg}
C.~Cordova, T.~T. Dumitrescu, and K.~Intriligator, ``{Exploring 2-Group Global
  Symmetries},'' \href{http://dx.doi.org/10.1007/JHEP02(2019)184}{{\em JHEP}
  {\bfseries 02} (2019) 184},
\href{http://arxiv.org/abs/1802.04790}{{\ttfamily arXiv:1802.04790 [hep-th]}}.

\bibitem{Tanizaki:2019rbk}
Y.~Tanizaki and M.~Unsal, ``{Modified instanton sum in QCD and
  higher-groups},'' \href{http://dx.doi.org/10.1007/JHEP03(2020)123}{{\em JHEP}
  {\bfseries 03} (2020) 123},
\href{http://arxiv.org/abs/1912.01033}{{\ttfamily arXiv:1912.01033 [hep-th]}}.

\bibitem{Hidaka:2020iaz}
Y.~Hidaka, M.~Nitta, and R.~Yokokura, ``{Higher-form symmetries and 3-group in
  axion electrodynamics},''
  \href{http://dx.doi.org/10.1016/j.physletb.2020.135672}{{\em Phys. Lett. B}
  {\bfseries 808} (2020) 135672},
  \href{http://arxiv.org/abs/2006.12532}{{\ttfamily arXiv:2006.12532
  [hep-th]}}.

\bibitem{Hidaka:2020izy}
Y.~Hidaka, M.~Nitta, and R.~Yokokura, ``{Global 3-group symmetry and 't Hooft
  anomalies in axion electrodynamics},''
  \href{http://dx.doi.org/10.1007/JHEP01(2021)173}{{\em JHEP} {\bfseries 01}
  (2021) 173}, \href{http://arxiv.org/abs/2009.14368}{{\ttfamily
  arXiv:2009.14368 [hep-th]}}.

\bibitem{Kapustin:1999ha}
A.~Kapustin and M.~J. Strassler, ``{On mirror symmetry in three-dimensional
  Abelian gauge theories},''
  \href{http://dx.doi.org/10.1088/1126-6708/1999/04/021}{{\em JHEP} {\bfseries
  04} (1999) 021}, \href{http://arxiv.org/abs/hep-th/9902033}{{\ttfamily
  arXiv:hep-th/9902033}}.

\bibitem{Witten:2003ya}
E.~Witten, ``{SL(2,Z) action on three-dimensional conformal field theories with
  Abelian symmetry},'' in {\em {From Fields to Strings: Circumnavigating
  Theoretical Physics: A Conference in Tribute to Ian Kogan}}, pp.~1173--1200.
\newblock 7, 2003.
\newblock \href{http://arxiv.org/abs/hep-th/0307041}{{\ttfamily
  arXiv:hep-th/0307041}}.

\bibitem{Cardy:1981qy}
J.~L. Cardy and E.~Rabinovici, ``{Phase Structure of Z(p) Models in the
  Presence of a Theta Parameter},''
\href{http://dx.doi.org/10.1016/0550-3213(82)90463-1}{{\em Nucl. Phys.}
  {\bfseries B205} (1982) 1--16}.

\bibitem{Cardy:1981fd}
J.~L. Cardy, ``{Duality and the Theta Parameter in Abelian Lattice Models},''
\href{http://dx.doi.org/10.1016/0550-3213(82)90464-3}{{\em Nucl. Phys.}
  {\bfseries B205} (1982) 17--26}.

\bibitem{Honda:2020txe}
M.~Honda and Y.~Tanizaki, ``{Topological aspects of $4$D Abelian lattice gauge
  theories with the $\theta$ parameter},''
  \href{http://dx.doi.org/10.1007/JHEP12(2020)154}{{\em JHEP} {\bfseries 12}
  (2020) 154}, \href{http://arxiv.org/abs/2009.10183}{{\ttfamily
  arXiv:2009.10183 [hep-th]}}.

\bibitem{Hayashi:2022fkw}
Y.~Hayashi and Y.~Tanizaki, ``{Non-invertible self-duality defects of
  Cardy-Rabinovici model and mixed gravitational anomaly},''
  \href{http://dx.doi.org/10.1007/JHEP08(2022)036}{{\em JHEP} {\bfseries 08}
  (2022) 036}, \href{http://arxiv.org/abs/2204.07440}{{\ttfamily
  arXiv:2204.07440 [hep-th]}}.

\bibitem{Vafa:1994tf}
C.~Vafa and E.~Witten, ``{A Strong coupling test of S duality},''
  \href{http://dx.doi.org/10.1016/0550-3213(94)90097-3}{{\em Nucl. Phys. B}
  {\bfseries 431} (1994) 3--77},
  \href{http://arxiv.org/abs/hep-th/9408074}{{\ttfamily arXiv:hep-th/9408074}}.

\bibitem{Donagi:1995cf}
R.~Donagi and E.~Witten, ``{Supersymmetric Yang-Mills theory and integrable
  systems},'' \href{http://dx.doi.org/10.1016/0550-3213(95)00609-5}{{\em Nucl.
  Phys. B} {\bfseries 460} (1996) 299--334},
  \href{http://arxiv.org/abs/hep-th/9510101}{{\ttfamily arXiv:hep-th/9510101}}.

\bibitem{Dorey:1999sj}
N.~Dorey, ``{An Elliptic superpotential for softly broken N=4 supersymmetric
  Yang-Mills theory},''
  \href{http://dx.doi.org/10.1088/1126-6708/1999/07/021}{{\em JHEP} {\bfseries
  07} (1999) 021},
\href{http://arxiv.org/abs/hep-th/9906011}{{\ttfamily arXiv:hep-th/9906011
  [hep-th]}}.

\bibitem{Polchinski:2000uf}
J.~Polchinski and M.~J. Strassler, ``{The String dual of a confining
  four-dimensional gauge theory},''
  \href{http://arxiv.org/abs/hep-th/0003136}{{\ttfamily arXiv:hep-th/0003136}}.

\bibitem{Damia:2023ses}
J.~A. Damia, R.~Argurio, F.~Benini, S.~Benvenuti, C.~Copetti, and L.~Tizzano,
  ``{Non-invertible symmetries along 4d RG flows},''
  \href{http://arxiv.org/abs/2305.17084}{{\ttfamily arXiv:2305.17084
  [hep-th]}}.

\bibitem{PhysRevB.40.4709}
M.~den Nijs and K.~Rommelse, ``Preroughening transitions in crystal surfaces
  and valence-bond phases in quantum spin chains,''
  \href{http://dx.doi.org/10.1103/PhysRevB.40.4709}{{\em Phys. Rev. B}
  {\bfseries 40} (Sep, 1989) 4709--4734}.

\bibitem{Haldane:1983ru}
F.~D.~M. Haldane, ``{Nonlinear field theory of large spin Heisenberg
  antiferromagnets. Semiclassically quantized solitons of the one-dimensional
  easy Axis Neel state},''
\href{http://dx.doi.org/10.1103/PhysRevLett.50.1153}{{\em Phys. Rev. Lett.}
  {\bfseries 50} (1983) 1153--1156}.

\bibitem{Affleck:1987vf}
I.~Affleck, T.~Kennedy, E.~H. Lieb, and H.~Tasaki, ``{Rigorous Results on
  Valence Bond Ground States in Antiferromagnets},''
\href{http://dx.doi.org/10.1103/PhysRevLett.59.799}{{\em Phys. Rev. Lett.}
  {\bfseries 59} (1987) 799}.

\bibitem{kennedy1992hidden}
T.~Kennedy and H.~Tasaki, ``Hidden symmetry breaking and the haldane phase in
  s= 1 quantum spin chains,'' {\em Communications in mathematical physics}
  {\bfseries 147} (1992) 431--484.

\bibitem{Kennedy:1992ifl}
T.~Kennedy and H.~Tasaki, ``{Hidden Z2\texttimes{}Z2 symmetry breaking in
  Haldane-gap antiferromagnets},''
  \href{http://dx.doi.org/10.1103/PhysRevB.45.304}{{\em Phys. Rev. B}
  {\bfseries 45} no.~1, (1992) 304}.

\bibitem{oshikawa1992hidden}
M.~Oshikawa, ``{Hidden Z2\texttimes{}Z2 symmetry in quantum spin chains with
  arbitrary integer spin},'' {\em Journal of Physics: Condensed Matter}
  {\bfseries 4} no.~36, (1992) 7469.

\bibitem{Li:2023mmw}
L.~Li, M.~Oshikawa, and Y.~Zheng, ``{Non-Invertible Duality Transformation
  Between SPT and SSB Phases},''
  \href{http://arxiv.org/abs/2301.07899}{{\ttfamily arXiv:2301.07899
  [cond-mat.str-el]}}.

\end{thebibliography}\endgroup
\end{document}